\documentclass[letter, amsfonts, amssymb, amsmath, reprint, twoside, prb]{revtex4-1}
\usepackage[english]{babel}
\usepackage[utf8]{inputenc}

\usepackage{graphicx}
\usepackage{braket}
\usepackage{enumitem}

\usepackage{soul}
\usepackage{xcolor}
\usepackage{ulem}

\usepackage[caption=false]{subfig}

\newcommand{\mathcolorbox}[2]{\colorbox{#1}{$\displaystyle #2$}}

\newcommand{\sig}{\sigma} 
\newcommand{\bsig}{\bar{\sigma}}

\begin{document}
\title{{Exact solution of two simple non-equilibrium electron-phonon and electron-electron coupled systems: the atomic limit of the Holstein-Hubbard model and the generalized Hatsugai-Komoto model}}

\author{R.~D.~Nesselrodt and J.~K.~Freericks}
    \email[Correspondence email address: ]{rdn11@georgetown.edu}% Your name
    \affiliation{Department of Physics, Georgetown University,
              37th and O Sts. NW, Washington, D.C. 20057 USA
              %Tel.: 202-687-5984\\
              }

\date{\today} % Leave empty to omit a date

\begin{abstract}
One of the challenges in many-body physics is determining the effects of phonons on strongly correlated electrons. The difficulty arises from strong correlations at differing energy scales---for band metals, Migdal-Eliashberg theory accurately determines electron-phonon coupling effects due to the absence of vertex corrections---but strongly correlated electrons require a more complex description and the standard Migdal-Eliashberg approach does not necessarily apply. In this work, we solve for the atomic limit Green's function of the Holstein-Hubbard model with both time-dependent electron-electron and electron-phonon couplings.  {We then examine the photoemission spectra (PES) of this model in and out of equilibrium. Next we use similar methods to exactly solve an extended version of the Hatsugai-Komoto model, and examine its behavior in and out of equilibrium. These calculations lead us to propose using the first moment of the photoemission spectra to signal non-equilibrium changes in electron-electron and electron-phonon couplings.}

\end{abstract}

%\keywords{Holstein, Hubbard, electron, phonon,  atomic limit, time-dependent coupling, nonequilibrium,polaron, bipolaron}

\maketitle

\section{Introduction} \label{sec:intro}

Pump-probe spectroscopy is one of the most important tools for studying complex materials with strongly correlated degrees of freedom. The femtosecond time-resolution allows one to examine excitation and relaxation phenomena from different degrees of freedom (electronic, spin, lattice) at their natural timescales. The technique has been used to probe many different materials in the time-domain: the classification of charge-density-wave insulators \cite{CDWinsulators,CDWPumpProbe} and charge-stripe materials \cite{epcOscillations}; examining the charge gap of spin-density-wave materials \cite{SDWPumpProbe}; determining the electron-phonon coupling in a metal \cite{EPCinAmetal}; studying the response in high $T_c$ cuprates \cite{Tc1, Tc2,Tc3,PDHSSC} and in iron-based superconductors \cite{iron}. In a recent experimental study \cite{iron}, evidence of bosonic mode coupling was seen in LiFeAs. Other recent work \cite{2020eph} indicates that complex superconducting behavior cannot be understood in terms of electron-electron or electron-phonon coupling alone, but only by considering both effects. 

 Whether the pump pulse or the non-equilibrium state it generates could modify either the electron-phonon coupling or the electron-electron coupling in pump-probe experiments is an open question. The Shen group  demonstrated\cite{EPCEvidence} that the electron-phonon coupling in the high $T_c$ cuprates is related to a kink in the electronic dispersion near the Fermi energy. More recent work concludes that the strength of the kink is determined by a combination of electron-phonon and electron-electron coupling \cite{huang2020KINK}. A softening of the kink has been observed in pump-probe experiments\cite{EPCquenching}, but such a softening does not necessitate a dynamic change in the electron-phonon coupling due to nonequilibrium effects \cite{JimLexKink}. A recent experimental study of electron thermalization in laser-excited nickel\cite{RecentNickelPaper} suggests a reduction of electron repulsion may explain the observation of an increasing spectral red shift with increasing electron temperature. In Ref.~\onlinecite{Zhang8788} the authors claim to manipulate electron temperatures, and consequently, the electron-phonon coupling by varying the pump fluence.

The ability to directly extract the electron-phonon coupling from experiment is an area of active interest that has seen exciting recent developments. Namely, a new analysis technique called frequency-domain angle-resolved photoemission spectroscopy (FD-ARPES) enables observation of the electron-phonon coupling directly from experimental data \cite{DirectEPC}, and has been used to measure the electron-phonon coupling during a photo-induced insulator to metal transition \cite{DirectEPCexperiment}.

In this work, we examine the atomic limit of the Hubbard-Holstein (HH) model and allow for arbitrary changes in both the electron-electron and electron-phonon coupling (but do not determine their time-dependence from a self-consistent reaction to the pump pulse). Similar studies have been done previously \cite{Werner_2013,Werner_2015} for the lattice model,  but we are not aware of work that {solves the Green's function exactly, and allows for arbitrary functional forms for the couplings and simultaneous variations of} both electron-electron and electron-phonon couplings as functions of time.{ We also propose a non-equilibrium extension to the Hatsugai-Komoto (HK) model},\cite{HKoriginal} {which includes coupling to a static, zero momentum phonon whose solution is identical to the model considered here, save for a simple substitution. Then, by considering phenomenological coupling changes in both models, we propose a measure, called the ``first photoemission spectra (PES) moment", which allows us to track changes in subsystem couplings in ultrafast experiments.} We also note that the single and two particle atomic Green's functions presented here are the required inputs to a systematic strong-coupling-expansion on the lattice, generalizing previous work \cite{JimSCE}, which applied only to the Hubbard model.
Recently,\cite{SCE2020} a strong-coupling-expansion of the Holstein-Hubbard model with constant coupling parameters found a rich landscape of phases, including distinct pair-density-waves, spin-density-waves, charge-density-waves, and superconducting regions as well as regions of phase separation. 

Another class of materials that this theory might be relevant for are those that have been observed to have polarons (or possibly bipolarons) generated by nonequilibrium pumps. It is difficult to experimentally differentiate between polarons and bipolarons, so some of these systems may be creating bipolarons, which naturally occur in the atomic limit and often are strongly localized. In Ref.~\onlinecite{Polaron1}, the authors use pump-probe spectroscopy to measure polaron lifetimes in a mixed-valence perovskite material under high pressures. They also argue that photoexcitation is analogous to applying a high pressure to this system, evidenced by the response of the phonons and the fact that they both lead to an intervalence charge transfer. More recently, experiments show a photoinduced transition to a long-lived polaronic state in lead halide perovskite \cite{Polaron2}, and in Ref.~\onlinecite{Polaron3} the authors demonstrate a technique for qualitatively estimating the magnitude and shape of a photoinduced polaronic distortion via femtosecond diffuse x-ray scattering techniques. In Ref.~\onlinecite{chen2021temperaturedriven} the authors claim to observe dynamical formation of small polarons in a MoTe$_{2}$ Weyl semimetal via terahertz spectroscopy.

In this work, we provide an exact (equilibrium and non-equilibrium) solution to the atomic Hubbard-Holstein (HH) model for arbitrary electron-electron and electron-phonon couplings. In addition to serving as the starting point for strong-coupling-based perturbation theories, it also provides an important benchmark for non-equilibrium calculations, where exact results are rare. The reason this problem can be solved exactly lies in the fact that the electron number for each spin is conserved in the atomic limit. But, even with that, the dynamics is sufficiently complex that one needs to use advanced techniques to factorize the evolution operator in order to determine the exact results presented here. For strongly coupled systems, and particularly those above the renormalized Fermi temperature, we believe this simple system will be representative of strongly coupled materials where charge fluctuations decrease as we approach the atomic limit.

The summary of the paper is as follows. In Sec.~\ref{sec:formalism}, we define the time-dependent Hamiltonian and provide general expressions for the spectral moments. In Sec.~\ref{sec:onepart}{ we present the single-particle Green's function both before and after taking the fermionic trace, and discuss it's analytic agreement with the spectral moments.} In Sec.~\ref{sec:atomicPES} {we consider the result of changing couplings on observable quantities in the local Holstein-Hubbard model, and in Sec.}~\ref{sec:HKmodel} {we do the same for the modified HK model.}
Our conclusions are presented in Sec.~\ref{sec:conclusion}. Appendices summarize some of the more cumbersome formulas {as well as provide additional details about the derivations and enumerate the expectation values which are necessary to compare against the spectral moments. The two-particle Green's function, the other piece needed to calculate a strong-coupling-type expansion, is presented in the appendix, and the derivation verifying the agreement of the single-particle Green's function with the spectral moment sum rules up to third order is given in the supplemental material \cite{sup}. }
\section{Formalism} \label{sec:formalism}
 The Holstein-Hubbard model is the simplest model Hamiltonian which describes both electron-electron and electron-phonon interactions \cite{Hubbard,Holstein}. It includes electrons that hop from one lattice site to another and Einstein phonons with the same frequency on each site. The electrons interact with the phonons via a direct coupling of the charge to the phonon coordinate. The electrons also repel each other via an on-site Coulomb repulsion. In the atomic limit, the electrons do not hop. This means we cannot represent electric-field effects with vector potentials and these systems never have electric current flow. The Holstein-Hubbard model for the atomic site then becomes 
 \begin{equation}
 \label{HHham}
    \hat{\mathcal H}(t)=\hbar\omega\left (\hat{a}^\dagger\hat{a}+\frac12\right )+\big(g(t)\hat{x}-\mu\big)(\hat{n}_{\uparrow}+\hat{n}_{\downarrow})+U(t)\hat{n}_{\uparrow}\hat{n}_{\downarrow}
\end{equation}
where $\hat{n}_{\sigma}=\hat{c}^{\dagger}_{\sigma}\hat{c}_{\sigma}$ is the fermionic number operator, $\hat{c}_{\sigma}^{\dagger}~(\hat{c}_{\sigma})$ creates (destroys) an electron with spin $\sigma$, $\mu$ is the chemical potential, $\hat{a}^{\dagger}~(\hat{a})$ are the bosonic creation (annihilation operators), $\hat{x}=\sqrt{\frac{\hbar}{2m\omega}}(\hat{a}+\hat{a}^{\dagger})$ is the phonon coordinate operator (with $\omega$ the phonon frequency), $g(t)$ is the electron-phonon coupling, and $U(t)$ is the electron-electron coupling. We drop the zero point energy from the Hamiltonian because it just adds an overall constant shift to all energies and we set $\hbar=1$ for the remainder of this work. Note that because $[\hat{\mathcal H}(t),\hat{n}_{\sigma}]=0$, the total electron number (and the electron number for each spin) is conserved. This symmetry is what allows us to derive an exact expression for the retarded Green's function; it is immediately broken once hopping between lattice sites is introduced. Note that we start our system in equilibrium in the limit as $t\to-\infty$ and all of the time dependence in the coupling constants starts at some finite time after $t_{min}$.

The retarded Green's function is defined as 
\begin{equation}
\label{gfdef}\begin{aligned}
    g^R_{\sigma}(t_1,t_2)&=\mathcolorbox{white}{\theta(t_1-t_2)\left(g^>_{\sigma}(t_1,t_2)-g^<(t_1,t_2)\right)}\\
    &=-i\theta(t_1-t_2)\big\langle \big\{\hat{c}_{\sigma}(t_1),\hat{c}_{\sigma}^{\dagger}(t_2)\big\}_+\rangle 
    \end{aligned}
\end{equation}
where $\theta(t)$ is
the Heaviside function, $\langle \dots\rangle$ indicates the thermal average at the initial time $\frac{1}{\mathcal Z}\text{Tr}\big\{e^{-\beta\hat{\mathcal H}(t_{min})} \dots\big\}$, {and $g^>_{\sigma}(t_1,t_2),g^<_{\sigma}(t_1,t_2)$ are the greater and lesser Green's functions, given by }\begin{equation}
    \label{ggr}
  \mathcolorbox{white}{  g^>_{\sigma}(t_1,t_2)=-i\langle \hat{c}_{\sigma}(t_1)\hat{c}_{\sigma}^{\dagger}(t_2)\rangle}
\end{equation}
and
\begin{equation}
  \mathcolorbox{white}{  \label{gles}
    g^<_{\sigma}(t_1,t_2)=i\langle \hat{c}^{\dagger}_{\sigma}(t_2)\hat{c}_{\sigma}(t_1)\rangle.}
\end{equation}
All time-dependence of operators is in the Heisenberg representation,
\begin{equation} \hat{O}_H(t)=\hat{U}^{\dagger}(t,t_{min})\hat{O}\hat{U}(t,t_{min})\end{equation} with $\hat{U}(t,t_{min})$ the time-evolution operator, \begin{equation}
    \hat{U}(t,t_{min}) = \mathcal{T}_t\exp\left[-i\int_{t_{min}}^t\!\!\!dt'\ \hat{\mathcal H}(t')\right].
\end{equation}
Here, $\mathcal{T}_t$ is the time-ordering operator, which places the latest times to the left and the Hamiltonian is in the Schr\" odinger representation; $t_{min}$ is a time where the system is in thermal equilibrium and is before any coupling constants start varying with time. 
The spectral moments of the Green's function are defined via
\begin{equation}
    \mu^{Rn}_{\sigma}(t_{ave})=-\frac{1}{\pi}\int_{-\infty}^{\infty}\,d\omega\ \omega^n\text{Im}\,g^R_{\sigma}(t_{ave},\omega)
\end{equation}
where 
\begin{eqnarray}
    &~&g_{\sigma}^R(t_{ave},\omega)=\nonumber\\
    &~&~~\int_0^{\infty}\,dt_{rel}\ e^{i\omega t_{rel}}g^R_{\sigma}\left (t_{ave}+\frac{t_{rel}}{2},t_{ave}-\frac{t_{rel}}{2}\right )
\end{eqnarray}
and we are now using Wigner coordinates of average and relative time,  defined by $t_{ave}=\frac{t_{1}+t_{2}}{2}$ and $t_{rel}=t_1-t_2$. The moments can be rewritten as derivatives with respect to relative time via
\begin{align}
\label{derivs}
    &\mu_{\sigma}^{Rn}(t_{ave})=\lim_{t_{rel}\to 0}\\
    &~~~\text{Re}\bigg\langle i^{n}\frac{d^n}{dt^n_{rel}}\left\{\hat{c}_{\sigma}\left (t_{ave}+\frac{t_{rel}}{2}\right ),\hat{c}_{\sigma}^{\dagger}\left (t_{ave}-\frac{t_{rel}}{2}\right )\right\}_+\bigg\rangle.\nonumber
\end{align}

{Finally, the time-resolved photoemission spectrum (tr-PES) is determined via the lesser Green's function (when we neglect matrix-element effects) by }\cite{Freericks_2009}\begin{widetext}\begin{equation}
   \mathcolorbox{white}{ P(t_{ave},\omega)=-i\int_{t_{min}}^{t_{ave}}\,dt_1\ \int_{t_{min}}^{t_{ave}}\,dt_2\ s(t_1)s(t_2)e^{i\omega(t_1-t_2)}g^<_{\sigma}(t_1,t_2)}
\end{equation}
\end{widetext}
{where $s(t)$ is the probe-pulse envelope, given by} \begin{equation}
    \mathcolorbox{white}{s(t)=\frac{1}{\Gamma\sqrt{\pi}}e^{-\frac{(t-t_0)^2}{\Gamma^2}}}.
\end{equation}
{Here $\Gamma$ is the effective temporal probe width, and $t_0$ is the time delay with respect to the application of the driving field pulse}.

\section{Holstein-Hubbard Green's Function}
\label{sec:onepart}
To evaluate the single-particle Green's function in Eq.~(\ref{gfdef}), we must take a trace over the Hilbert space composed of a direct product of the Fermionic states with the simple harmonic oscillator states.
Newton's generalized binomial theorem enables us to sum the infinite degrees of freedom of the harmonic oscillator to obtain a closed form expression for the Bosonic trace. This follows the technique used in Ref.~\onlinecite{jimHolstein} to evaluate the partition function. We outline this procedure briefly here, as it is used repeatedly in our derivations.  We have 
\begin{align}
    g^R_{\sigma}(t_1,t_2)&=-\frac{i\theta(t_1-t_2)}{\mathcal Z}\nonumber\\
    &\times\text{Tr}_{b,f}\bigg\{e^{-\beta\hat{\mathcal H}(t_{min})}\{\hat{c}_{\sigma}(t_1),\hat{c}_{\sigma}^{\dagger}(t_2)\}_+\bigg\}
\end{align}
The Bosonic trace is calculated in the following way. First, we note that the single Bosonic operators always appear as summed pairs, which can be re-expressed by employing the Weyl form of the Baker-Campbell-Hausdorff identity as
\begin{equation}
e^{\alpha \hat{a}+\beta \hat{a}^{\dagger}}=e^{\alpha \hat{a}}e^{\beta \hat{a}^{\dagger}}e^{-\frac{1}{2}\alpha\beta[\hat{a},\hat{a}^{\dagger}]}.
\end{equation} 
Using this expression, we can reduce the Bosonic trace to  constants (with respect to the Bosonic operators) multiplied by the form 
\begin{equation}
    \text{Tr}_b\left\{e^{-\beta\omega \hat{a}^{\dagger}\hat{a}}e^{A\hat{a}}e^{B\hat{a}^{\dagger}}\right\},
\end{equation}
where $A$ and $B$ are constants \textit{with respect to the Bosonic operators}. Note that the simple harmonic oscillator states $\ket{n}$ are eigenvectors of the Bosonic number operator $\hat{n}=\hat{a}^{\dagger}\hat{a}$ with eigenvalue $n$, so we have 
\begin{align}
     &\text{Tr}_b\left\{e^{-\beta\omega \hat{a}^{\dagger}\hat{a}}e^{A\hat{a}}e^{B\hat{a}^{\dagger}}\right\}=\sum_{n=0}^{\infty}e^{-\beta\omega  n}\langle n|e^{A\hat{a}}e^{B\hat{a}^{\dagger}}|n\rangle.
\end{align}
By expanding each exponential inside the expectation value in a power series, using the properties of the simple harmonic oscillator states and summing the resulting series with Newton's generalized binomial theorem, we evaluate the trace with the result
\begin{align}
     &\text{Tr}_b\big\{e^{-\beta\omega \hat{a}^{\dagger}\hat{a}}e^{A\hat{a}}e^{B\hat{a}^{\dagger}}\big\}\nonumber\\
     &~~~~~=\frac{1}{1-e^{-\beta\omega}}\exp\left[\frac{1}{1-e^{-\beta\omega}}AB\right].
\end{align}

Using this result allows us to evaluate the Bosonic trace in the definition of the Green's function, which becomes
\begin{align}
\label{bosetrace}
    &g_{\sigma}^R(t_1,t_2)=\frac{-i\theta(t_1-t_2)\left(n(\omega)+1\right)}{\mathcal Z}e^{i\mu(t_1-t_2)}\nonumber\\ &~~~\times\exp\left[\left(-\frac{1}{2}-n(\omega)\right)\big|C(t_1)-C(t_2)\big|^2\right]\nonumber\\ &~~~\times
    \text{Tr}_f\bigg\{\exp\left[\frac{\beta g^2(t_{min})\hat{n}_f^2}{2m\omega^2}+\beta\mu \hat{n}_f-\beta U(t_{min})\hat{n}_{\uparrow}\hat{n}_{\downarrow}\right]\nonumber\\
    &~~~\times\exp\left[-i\int_{t_2}^{t_1}\,dt\ U(t)\hat{n}_{\bar{\sigma}}\right]\nonumber\\
    &~~~\times\exp\left[i(1+2\hat{n}_{\bar{\sigma}})\big(I(t_1)-I(t_2)\big)]\right]\nonumber\\
   &~~~\times \exp\left[\frac{2ig(t_{min})}{\sqrt{2m\omega^3}}\text{Re}\big\{C(t_1)-C(t_2)\big\}\hat{n}_f\right]\nonumber\\
    &~~~\times\left\{e^{i\text{Im}\{C^{*}(t_1)C(t_2)\}}\hat{c}_{\sigma}\hat{c}_{\sigma}^{\dagger}+ e^{-i\text{Im}\{C^{*}(t_1)C(t_2)\}}\hat{c}_{\sigma}^{\dagger}\hat{c}_{\sigma}\right\}\bigg\}
\end{align}
where
\begin{equation}
\label{Coft}
    C(t)=\frac{1}{\sqrt{2m\omega}}\int_{t_{min}}^t\!\!\!dt'\ g(t') e^{i\omega \left(t'-t_{min}\right)},
\end{equation}
and \begin{align}
\label{Ioft}
    I(t)=\frac{1}{2m\omega}\int_{t_{min}}^t\!\!\!dt' \int_{t_{min}}^{t'}\!\!\!dt''\ g(t')g(t'')\sin\omega(t'-t'')
\end{align}
are functions we define to make this result more readable, the partition function is given by \begin{align}
\label{partition}
    \mathcal{Z}&=\Big(n(\omega)+1\Big)\left\{1+2e^{\beta\mu}\exp\left[\frac{\beta g^2(t_{min})}{2m\omega^2}\right]\right.\nonumber\\
    &+\left .\exp\left[\beta(2\mu-U(t_{min}))\right]\exp\left[\frac{2\beta g^2(t_{min})}{m\omega^2}\right]\right\}
\end{align}
with the Bose-Einstein distribution function given by
\begin{align}
\label{bose}
    n(\omega)=\frac{1}{e^{\beta\omega}-1}.
\end{align}
We can now evaluate the Fermionic trace over the four possible states, with the result
\begin{widetext}
\begin{align}
\label{final1}
g_{\sigma}^R(t_1,t_2)&=-\frac{i\theta(t_1-t_2)\big (n(\omega)+1\big)}{\mathcal Z}e^{i\mu(t_1-t_2)}
\exp\bigg[\left(-\frac{1}{2}-n(\omega)\right)\big|C(t_1)-C(t_2)\big|^2\bigg]\nonumber\\&\times
\bigg\{e^{i\text{Im}\{C^{*}(t_1)C(t_2)\}}\bigg(\exp\big[i\big(I(t_1)-I(t_2)\big)\big]\nonumber\\
&+\exp\left[\beta(\mu+\frac{g^2(t_{min})}{2m\omega^2})\right] \exp\left[-i\int_{t_2}^{t_1}\,dt'\ U(t')\right]\exp\big[3i(I(t_1)-I(t_2))\big]
\exp\left[\frac{2ig(t_{min})}{\sqrt{2m\omega^3}}\text{Re}\{C(t_1)-C(t_2)\}\right ]\bigg)\nonumber\\
&+e^{-i\text{Im}\{C^{*}(t_1)C(t_2)\}}\bigg( \exp\left[\mathcolorbox{white}{\beta(\mu+\frac{g^2(t_{min})}{2m\omega^2}})\right] \exp\big[i\big(I(t_1)-I(t_2)\big)\big] \exp\left[\frac{2ig(t_{min})}{\sqrt{2m\omega^3}}\text{Re}\{C(t_1)-C(t_2)\}\right]\nonumber\\
&+\exp\left[\frac{2\beta g^2(t_{min})}{m\omega^2}+\beta(2\mu-U(t_{min}))\right]\exp\left[-i\int_{t_2}^{t_1}\,dt\ U(t)\right]\exp\big[3i(I(t_1)-I(t_2))\big]\nonumber\\
&\times\exp\left[\frac{4ig(t_{min})}{\sqrt{2m\omega^3}}\text{Re}\{C(t_1)-C(t_2)\}\right]\bigg)\bigg\}.
\end{align}
\end{widetext}

This result is quite complicated. But note that it is an exact expression, so it can be employed to describe the response of Holstein-Hubbard systems with arbitrary time-dependent interactions. It is remarkable that such a complex system has such a compact exact expression for the nonequilibrium Green's functions. We reiterate again that there are two keys to allowing this to occur: first the number operators for each spin are conserved and do not depend on time and second, the commutator of the electron-phonon interaction with itself at two different times (in the interaction representation) is an operator that commutes with all remaining operators in the system. Nevertheless, the final result is not simple. It represents the complex dynamics of electron-phonon coupled systems in the atomic limit. In contrast, the atomic limit Green's functions for the Hubbard model are quite simple.

While it would be nice to have an independent check to verify that these results are correct, we are not aware of any way to do so for the full functions. But we do demonstrate in the appendix that this Green's function does satisfy the first four spectral moment sum rules analytically, via differentiation of the Green's function in the form of Eq.~(\ref{bosetrace}). The first three are evaluated analytically by taking derivatives of Eq.~(\ref{bosetrace}), before we take the Fermionic trace, as in Eq.~(\ref{derivs}). We also numerically check the first three moments to ensure we have taken the trace correctly in going from Eq.~(\ref{bosetrace}) to Eq.~(\ref{final1}) (for more information see Ref. \onlinecite{sonapaper}). This represents a quite stringent test of accuracy of the final results. {Indeed, these exact results were employed to benchmark the nonequilibrium Green's function moments and discovered some errors in previous work on this topic}\cite{sonapaper}.

\section{Results: Single-particle Photoemission Spectra}
\label{sec:atomicPES}
{Having obtained an exact result for an interacting, non-equilibrium system, we ask what occurs as a consequence of changing electron-electron and electron-phonon couplings. To this end we calculate the photoemission spectrum in and out of equilibrium. }
\begin{figure}[t]
\centering
\subfloat[fig1][Equilibrium]{
\includegraphics[width=0.45\columnwidth]{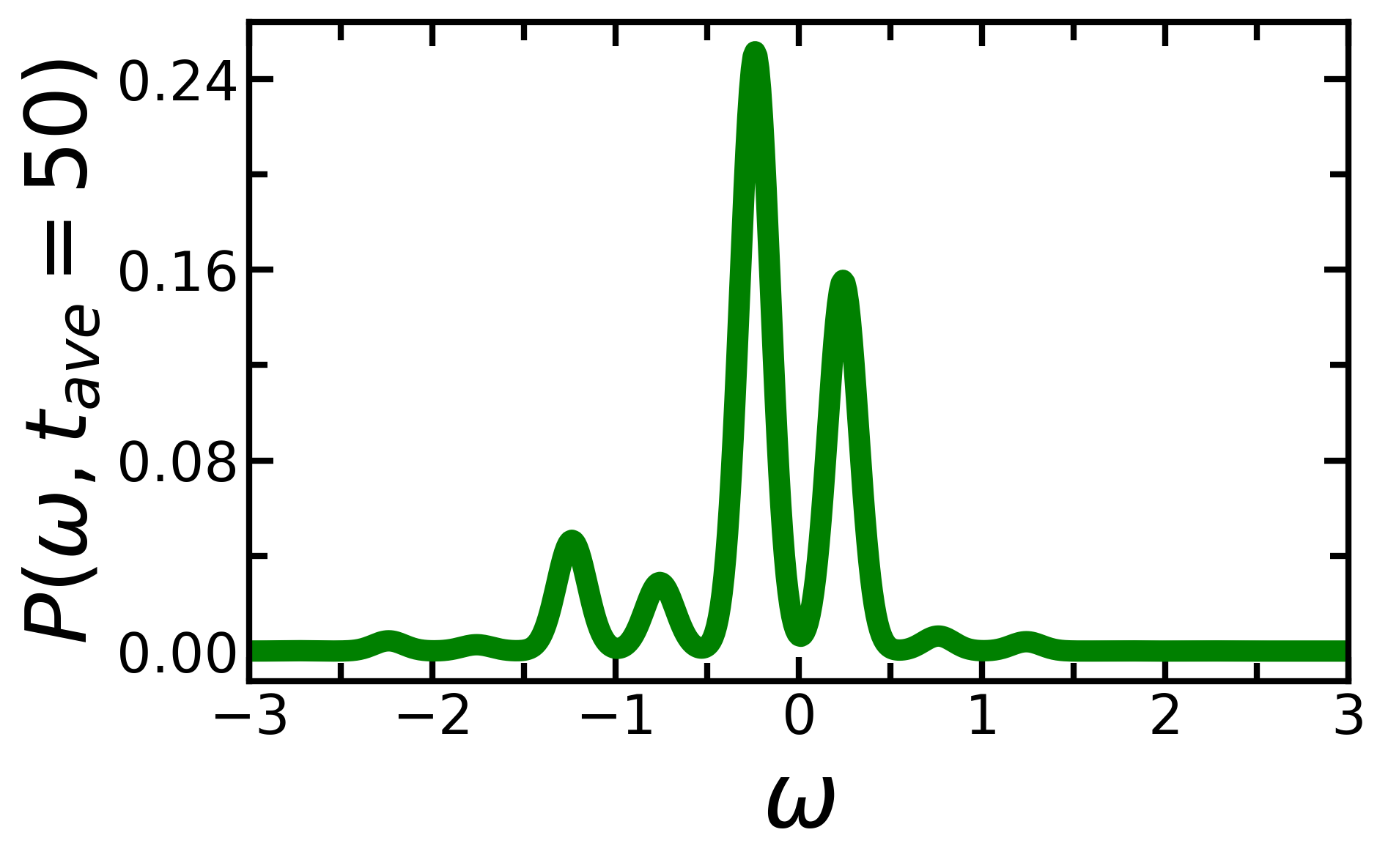}
\label{fig:1a}}
\subfloat[fig2][$g(t)$ decreasing 20\%]{
\includegraphics[width=0.45\columnwidth]{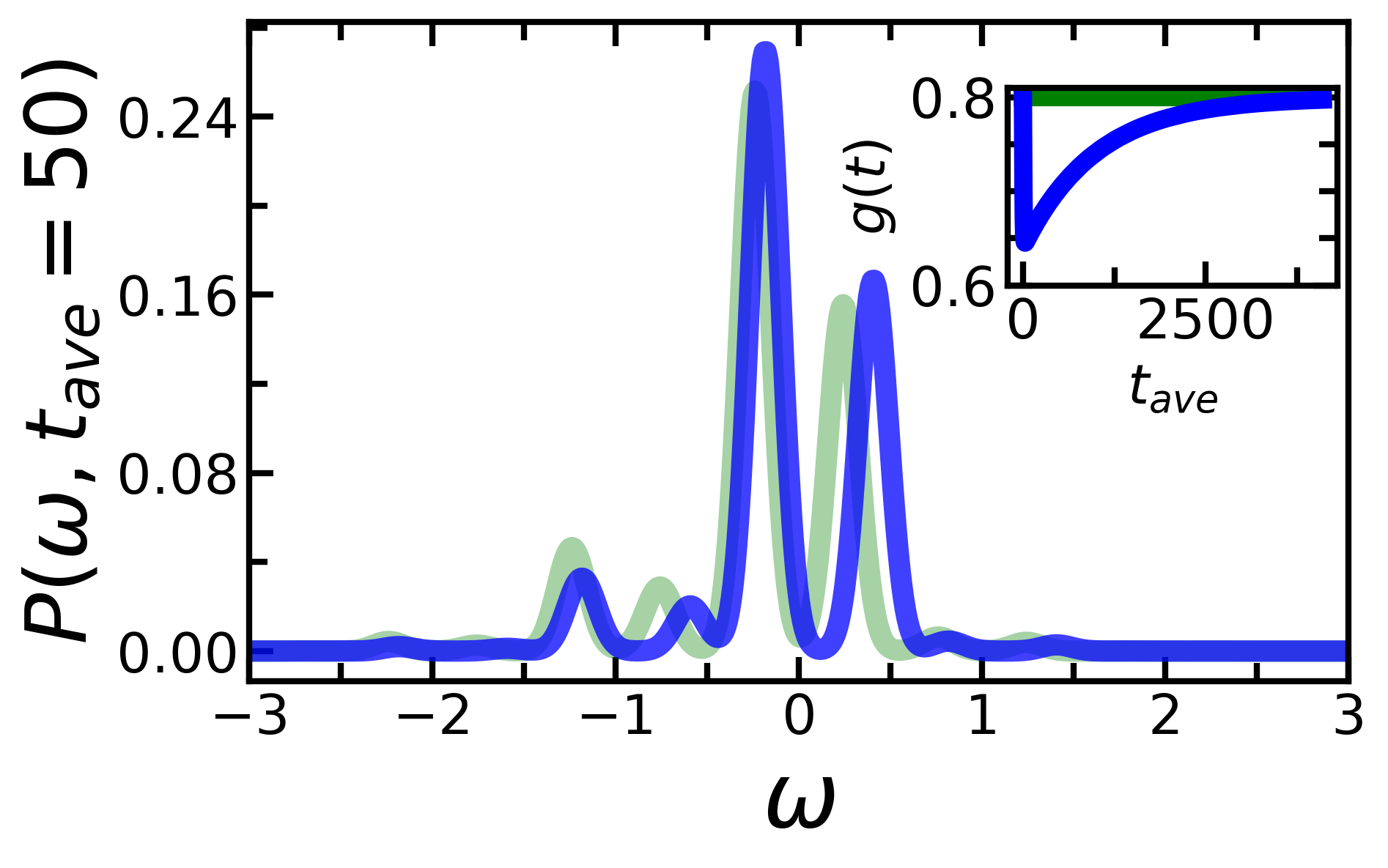}
\label{fig:1b}}
\qquad
\subfloat[fig3][$U(t)$ decreasing 20\%]{
\includegraphics[width=0.45\columnwidth]{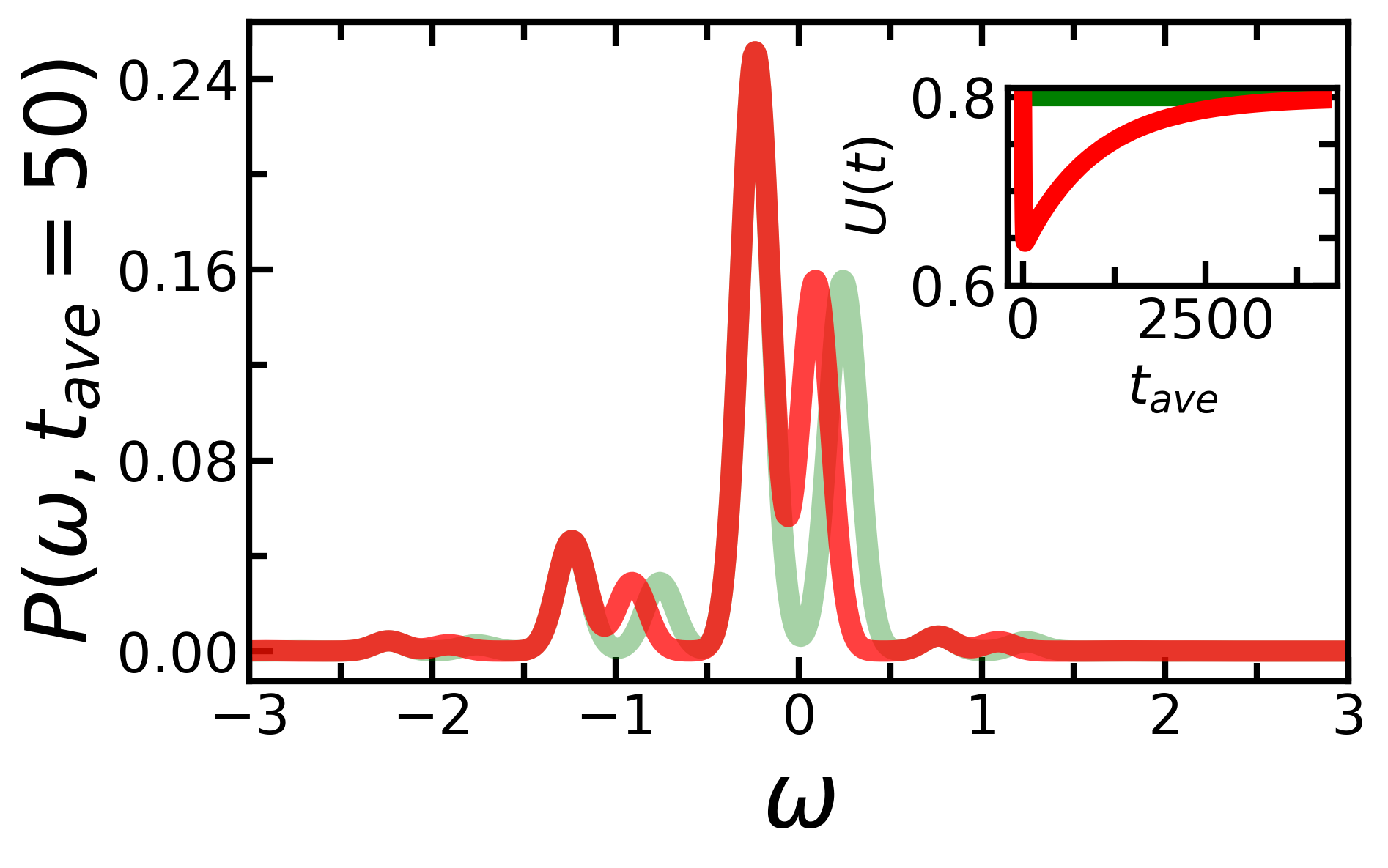}
\label{fig:1c}}
\subfloat[fig4][$g(t),U(t)$ both decreasing 20\%]{
\includegraphics[width=0.45\columnwidth]{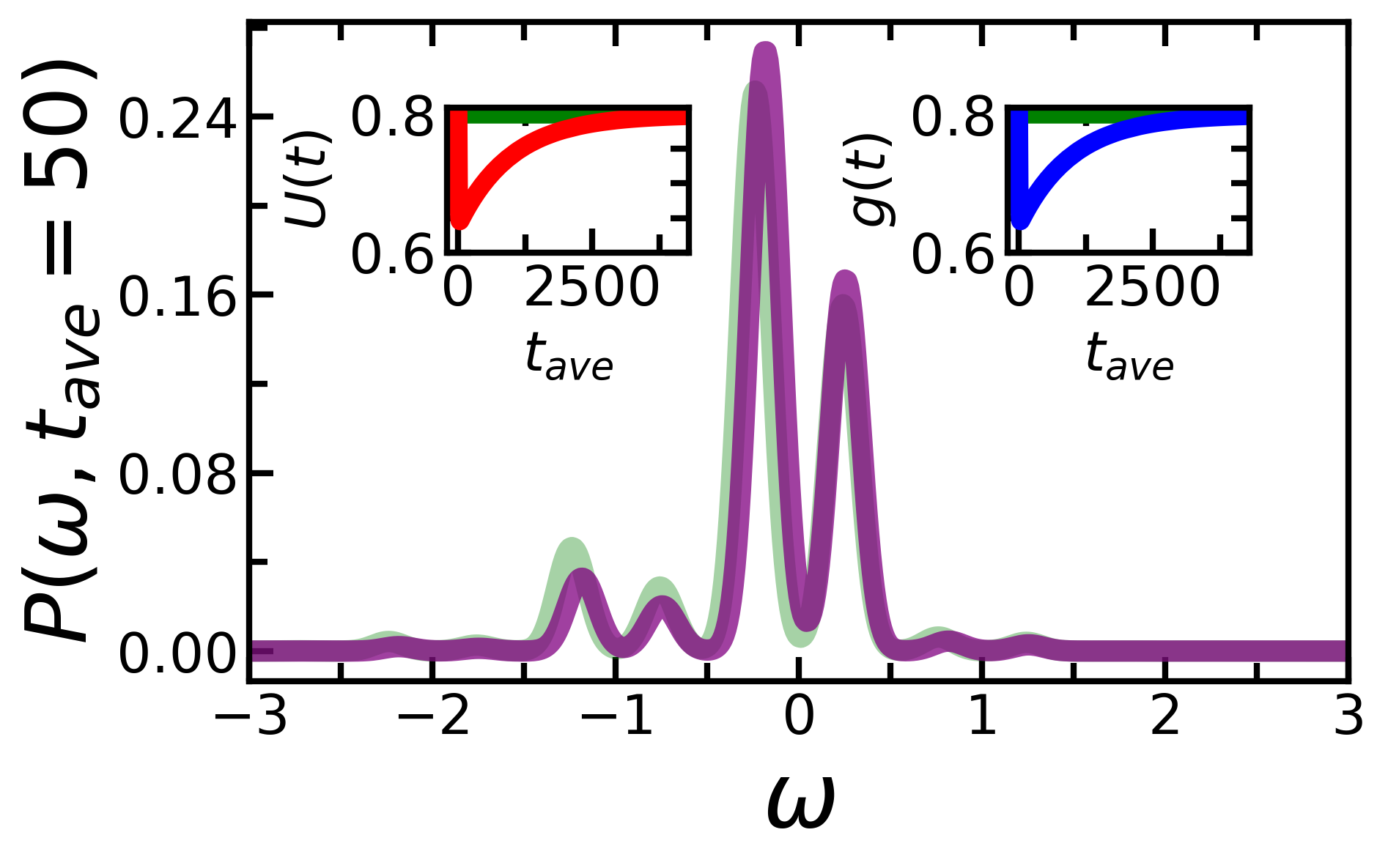}
\label{fig:1d}}
\caption{Demonstration of the effect of coupling changes on PES in the local HH model. Note that changes in e-ph coupling shift each peak (\ref{fig:1b}), while changes in e-e coupling shift every other peak (\ref{fig:1c}).}
\label{fig:1}
\end{figure}

{Motivating the phenomenological coupling changes presented here are numerous works studying the temperature dependence of the electron-phonon coupling in different materials} \cite{tempEPC1,EPCtemp2,TempEPC3,tempdepG,han2021raman} {as well as possible coupling changes induced by photoexcitation or the resulting nonequilibrium state} \cite{EPCinAmetal,epctemp4,failuretypicalEPC}.  Reference~\onlinecite{EPCinAmetal} {suggests that photoexcitation in a metal suppresses the electron-phonon coupling, which then gradually approaches its equilibrium value as a function of time. With this in mind, we measure the non-equilibrium photoemission spectra with varying couplings and compare to equilibrium.

In equilibrium, the photoemission spectra of the local HH model has several possible sets of peaks centered at $\omega=-n\frac{g^2}{2m\omega_0^2}-\mu$ and $\omega=-n\frac{g^2}{2m\omega_0^2}-\mu+U$ with $\omega_0$ the phonon frequency and $n\in\{1,2,3,4\}$. Each set of these peaks is spaced by $\omega_0$, and in regions of parameter space two sets of peaks can coexist, as in Fig}.~\ref{fig:1a}, {giving an overall spacing between the peaks of $\frac{1}{2}\omega_0$. For example, in} Fig.~\ref{fig:1a} {the main peak is centered at $\omega=-\frac{g^2}{2m\omega_0^2}-\mu$ and the secondary peak is centered at $\omega=-\frac{3g^2}{2m\omega^2}-\mu+U$. The envelope of the spectra has an inverse relationship to the inverse temperature $\beta$.} 

\begin{figure}[t]
\centering
\subfloat[][First PES moment for various e-ph coupling changes (given in inset) for the local HH model.]{\includegraphics[width=0.85\columnwidth]{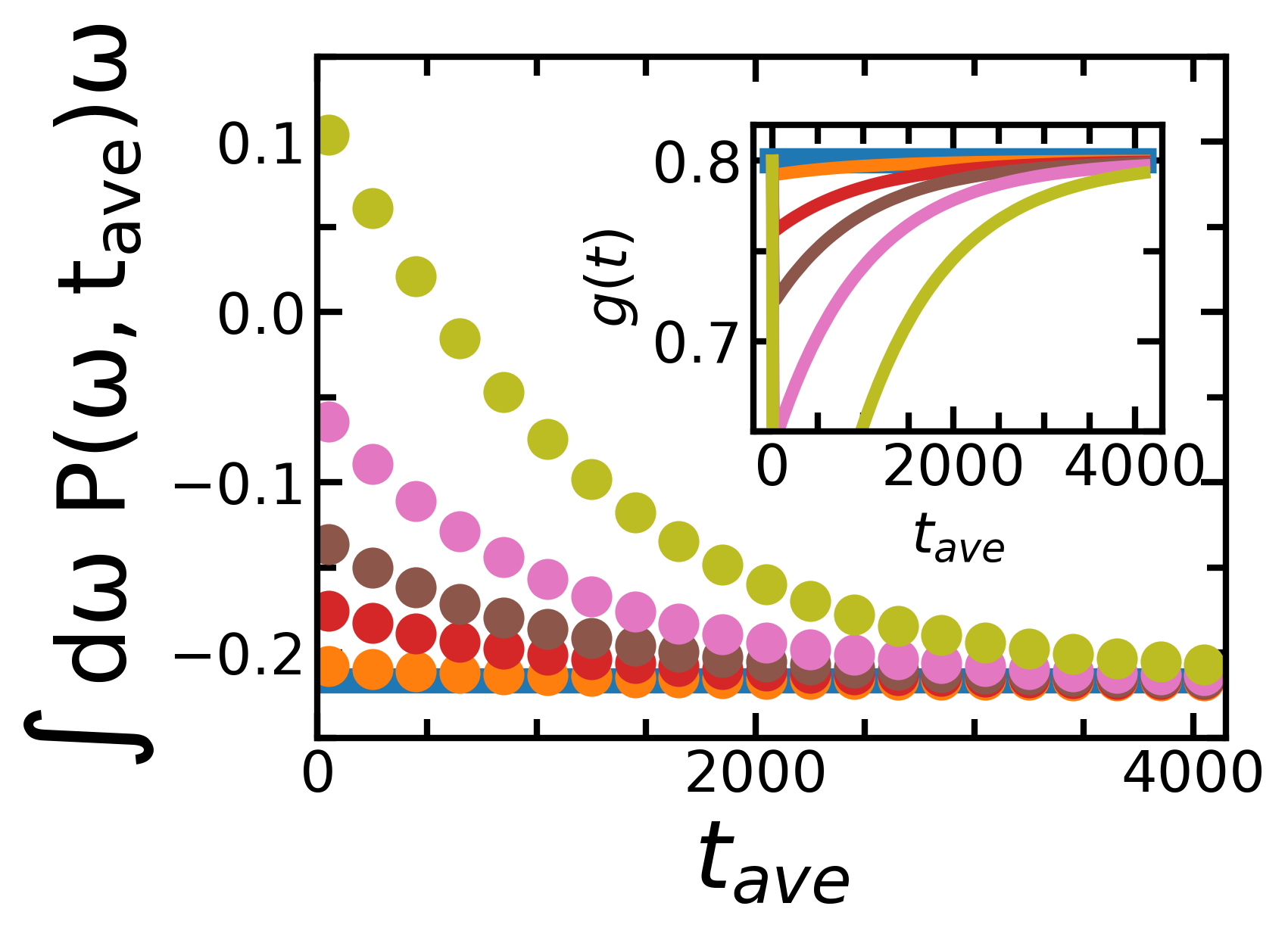}\label{fig:2a}}\par
\subfloat[][First PES moment for various e-e coupling changes (given in inset) for the local HH model.]{\includegraphics[width=0.85\columnwidth]{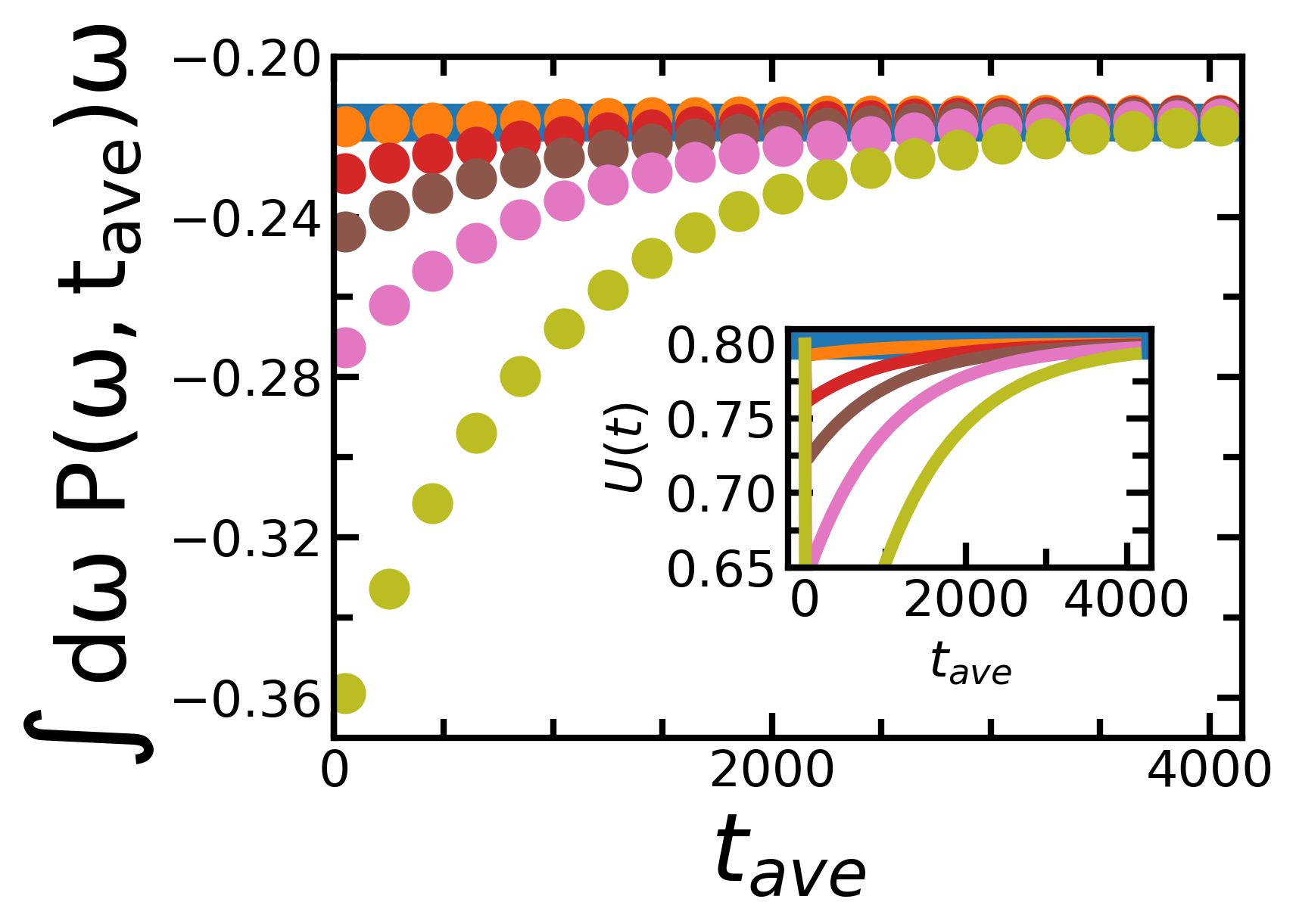}\label{fig:2b}}\par
\subfloat[][First PES moment for various combined e-ph and e-e coupling changes (given in inset) for the local HH model.]{ \includegraphics[width=0.85\columnwidth]{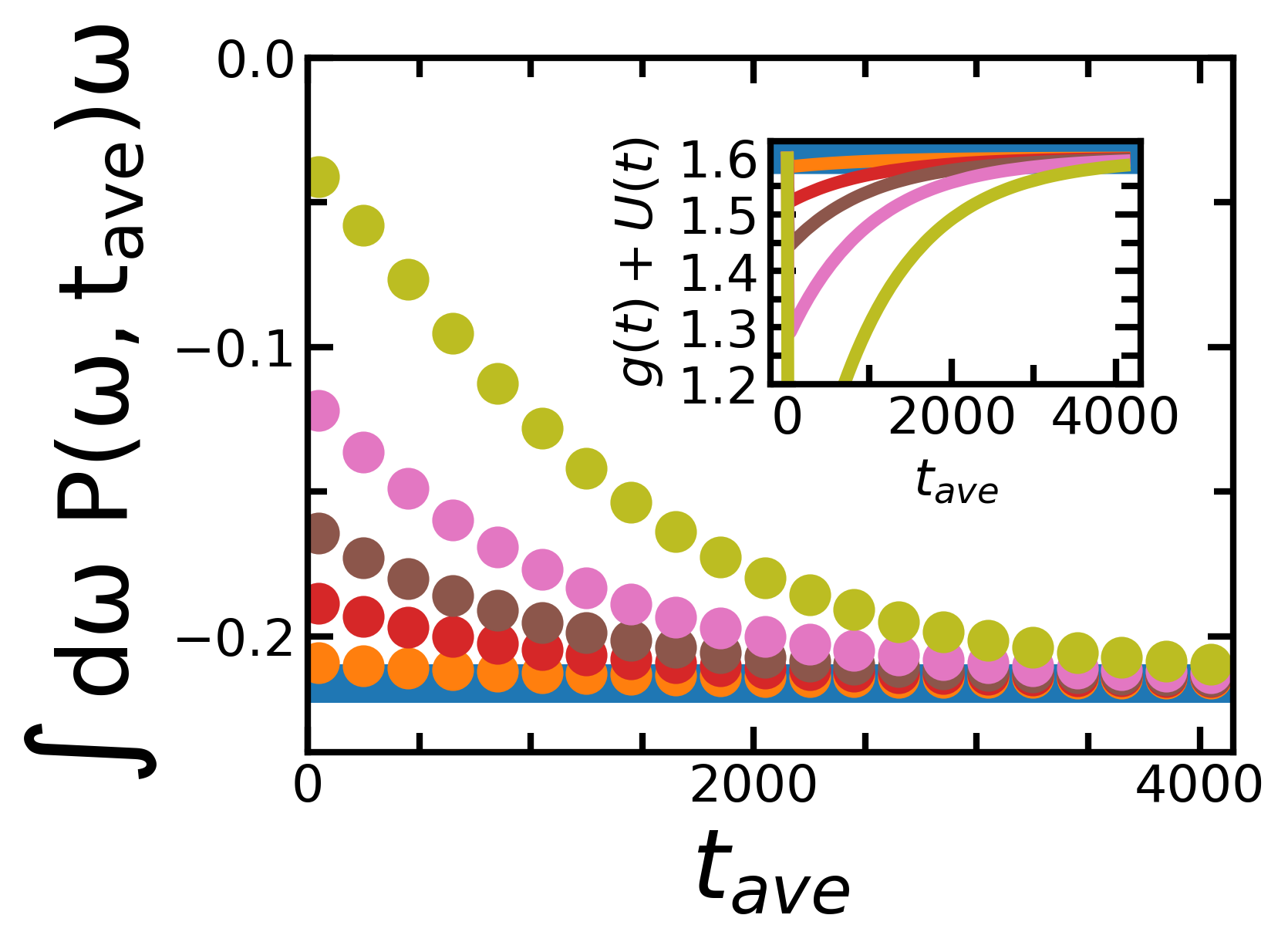}\label{fig:2c}}
\caption{First PES moment for the local HH model.}%
\label{fig:2}
\end{figure}

{As seen in} Fig.~\ref{fig:1},{ a decrease in electron-phonon coupling shifts the spectra to higher energies, while a decrease in electron-electron coupling shifts the spectra towards lower energies. This leads us to suggest a measure, called the first moment of the photoemission (or first PES moment), given by $\int\!d\omega\ \!\omega P(t_{ave},\omega)$, which shifts as a result of changes in the couplings. Note in Fig.}~\ref{fig:1} {that a changing electron-phonon coupling shifts every peak, while a changing electron-electron coupling shifts every other. This suggests that the peaks which are shifted by a changing electron-electron coupling are associated with double-electron occupancies, while the other peaks are associated with single-electron occupancy. }

{Shifts of spectral weight during photoexcitation have been observed experimentally} \cite{RecentNickelPaper,Germanium}. In Ref.~\onlinecite{RecentNickelPaper}, {the authors suggest a reduction in electronic repulsion as a possible explanation of the observed spectral redshift. Our proposed measure of the first PES moment could provide evidence for and quantify the degree of dynamical coupling changes in correlated systems.}

In Fig.~\ref{fig:2}, {we plot the first PES moment as a function of average time for couplings which fall continuously to $1,5,10,20$ and $50$ percent of their equilibrium values and exponentially recover, as plotted in the insets. In} Fig.~(\ref{fig:2a}), {we see that the first PES moment increases when the e-ph coupling decreases. We see the opposite behavior for the e-e case }(\ref{fig:2b}). {When both change simultaneously, the e-ph change dominates, as seen in Fig.}~(\ref{fig:2c}). {We feel that the first PES moment provides a nice, simple experimental measure that indicates changes in the system couplings for a system with sufficiently separated bands. In the next section, we demonstrate this behavior persists when extended to an exactly solvable non-equilibrium lattice model.}
\section{Application: Modified Hatsugai-Komoto Model}
\label{sec:HKmodel}
{Recently there has been a growing interest}\cite{Phillips_2020,HKelCorr,huang2021doped,HKvary} in a simple model Hamiltonian introduced by Hatsugai and Komoto in Refs.~\onlinecite{HKoriginal} and \onlinecite{bask},{ in part due to its potential as a solvable model of non-conventional superconductivity. Reference~\onlinecite{Phillips_2020}, {demonstrates that the Hatsugai-Komoto (HK) model possesses a Mott insulating phase and a ``strange metallic" phase which violates Luttinger's theorem and possesses a superconducting instability}  \cite{Phillips_2020}. {It also demonstrates that this model possesses a ``spectral weight transfer" similar to behavior observed in the cuprates}\cite{Molegraaf2239} {in the superconducting state, and that this superconducting state is distinct from the typical Bardeen-Cooper-Schrieffer (BCS) type superconducting state.}
\begin{figure}[t]
\centering
\subfloat[][First PES moment for various e-ph coupling changes (given in inset) for the HK$^{+}$ model.]{\includegraphics[width=0.8\columnwidth]{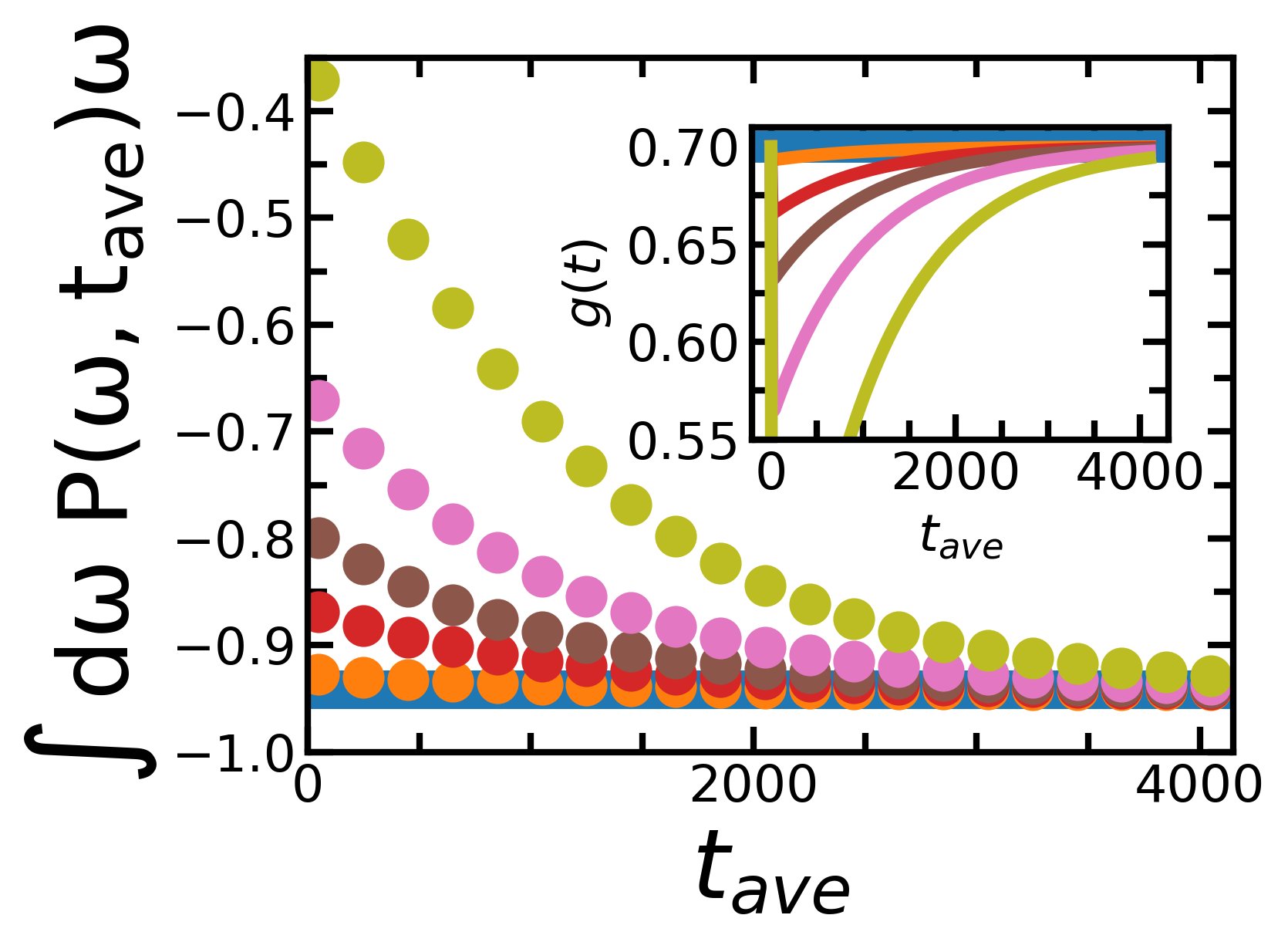}\label{fig:3a}}\par
\subfloat[][First PES moment for various e-e coupling changes (given in inset) for the HK$^{+}$ model.]{\includegraphics[width=0.8\columnwidth]{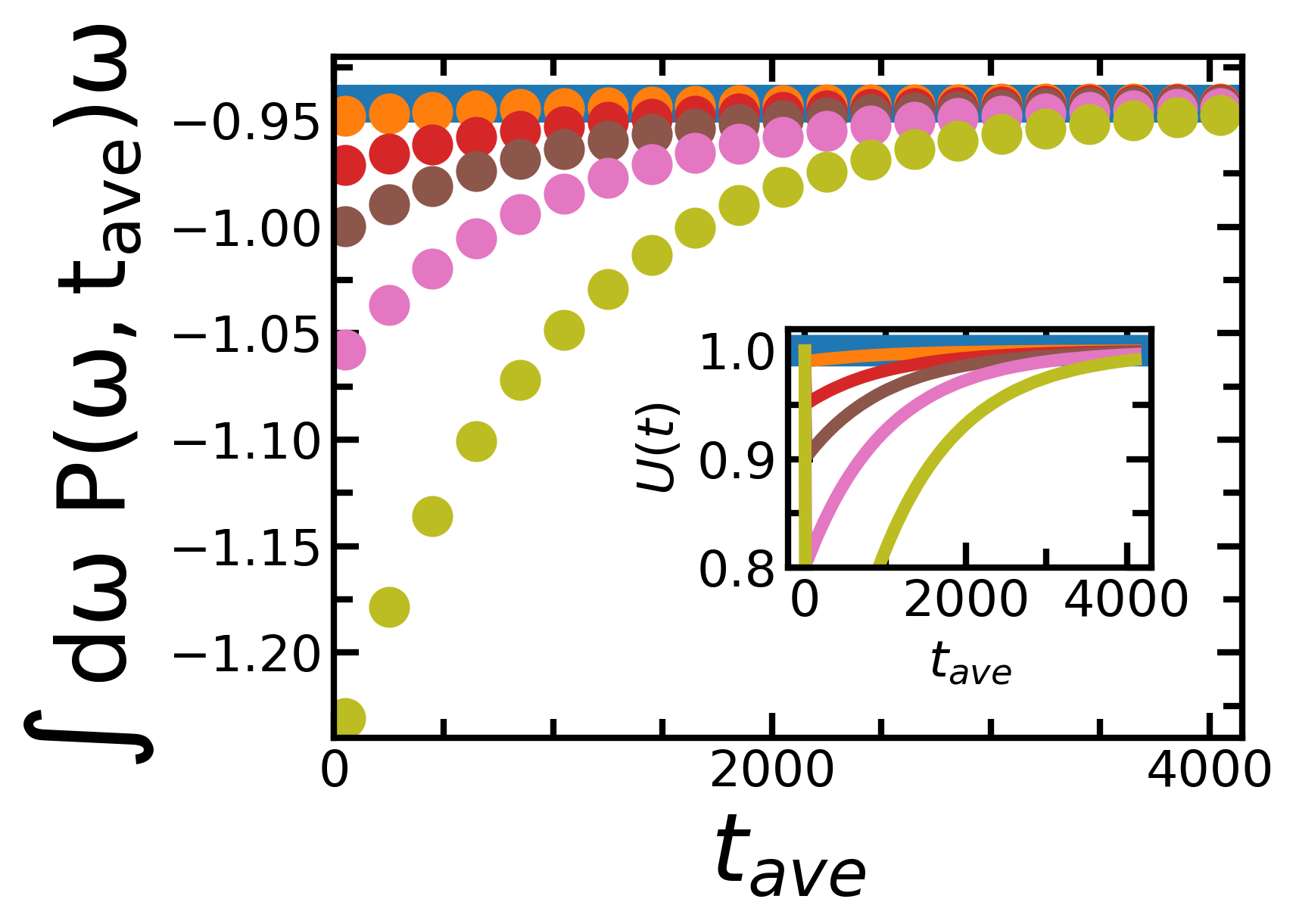}\label{fig:3b}}\par
\subfloat[][First PES moment for various combined e-ph and e-e coupling changes (given in inset) for the HK$^{+}$ model.]{ \includegraphics[width=0.8\columnwidth]{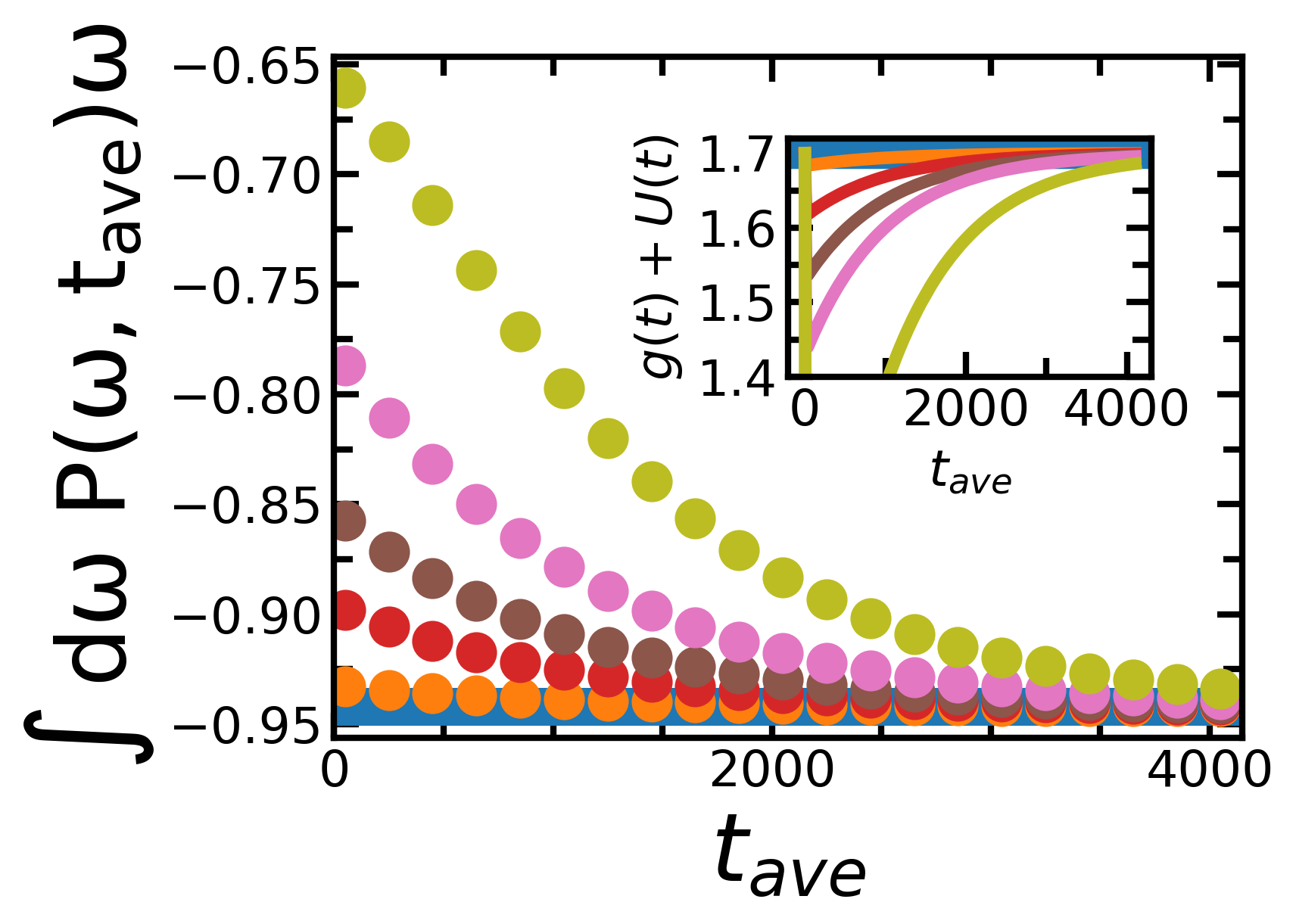}\label{fig:3c}}
\caption{First PES moment for the HK$^{+}$ model.}%
\label{fig:3}
\end{figure}

{We propose a non-equilibrium extension of this model, denoted $\hat{H}^{HK+}$ where the original HK model is given a time dependence $U(t)$ and is coupled to a zero momentum phonon via an electron-phonon coupling $g(t)$,}\begin{align}
    &\hat{H}^{HK+}(t)=\sum_{\boldsymbol{k}}\hat{H}_{\boldsymbol{k}}(t)\nonumber\\
    &=\sum_{\boldsymbol{k}}\bigg[\left(\epsilon_{\boldsymbol{k}}-\mu+g(t)\hat{x}_0\right)\hat{n}^f_{\boldsymbol{k}}\nonumber\\
    &+U(t)\hat{n}_{\boldsymbol{k}\sigma}\hat{n}_{\boldsymbol{k}\bar{\sigma}}+\delta_{\boldsymbol{k},0}\omega\left(\hat{a}^{\dagger}_{\boldsymbol{k}}\hat{a}_{\boldsymbol{k}}+\frac{1}{2}\right)\bigg] 
\end{align}
{where the sum is taken over the first Brillouin zone, $\boldsymbol{k} \in [-\pi,\pi)^d$ where $d$ the dimensionality, $\epsilon_{\boldsymbol{k}}$ is the band structure, and $\hat{x}_0=\frac{1}{\sqrt{2m\omega}}\left(\hat{a}_0^{\dagger}+\hat{a}_0\right)$ is the zero-momentum phonon coordinate. By comparison to} Eq.~\eqref{HHham}, we see this model is solved by Eq.~\eqref{final1} {with the substitution $\mu\to -\epsilon_{\boldsymbol{k}}+\mu$ giving the exact solution for each momentum point. Summing over the first Brillouin zone gives the exact solution of the full model. The dimensionality of the model enters via the kinetic energy term. For simplicity, we consider the one-dimensional case.}

{For small e-ph couplings this model behaves much like the original HK model, with a Mott insulating phase and a non-Fermi liquid phase. The presence of a non-zero electron-phonon coupling gives rise to peaks in the density of states and photoemission spectra spaced by the phonon frequency. These peaks tend to metallize the Mott phase for large enough couplings, giving rise to a ``near" Mott insulator with non-Fermi liquid behavior. As the e-ph coupling increases, the phonon peaks grow and completely destroy the psuedo-Mott phase.}

{In} Fig.~\ref{fig:3}, {we examine the first PES moment, as in the local HH model, in the non-fermi liquid metallic phase of the extended HK$^{+}$ model. Here we again see that decreases in the e-ph coupling cause an increase in the first PES moment, which outweighs the decrease caused by a reduction of e-e coupling. This demonstrates that shifts in the first PES moment as a result of coupling changes are a feature that appears in models more diverse than the local HH model. This suggests that these spectral moment shifts may be a general feature that can be used to indicate a change in couplings between various subsystems and be directly observed in ultrafast experiments.}

\section{Conclusion}
\label{sec:conclusion}
In this work we have found an exact result for the single-particle correlation function of the non-equilibrium Holstein-Hubbard model (with time dependent interactions) in the atomic limit. This includes all of the complex dynamical effects associated with the coupling of electrons to quantum phonons and the possible change in time of the electron-electron and electron-phonon interactions. Such dynamics are of course, quite complex, so the results are cumbersome. We verify these results by comparing to the exact results of the first four non-equilibrium spectral moments of the single-particle causal Green's function, finding exact analytic agreement. In Appendix \ref{sec:twopart} we calculate the two-particle Green's function for this model. Both of these Green's functions are needed to calculate the lattice Green's functions within a strong-coupling-expansion framework. We believe such future work will eventually  provide insight into the behavior of complex materials in pump-probe spectroscopy experiments, especially those which host bipolarons.

{We then address the question: if the couplings change as a function of time, what would be a possible experimental signature of these changes? To do this we pick a phenomenological function for the coupling changes, and find that the ``first PES moment," given by an integral of the frequency times the photoemission spectra, shifts with changes in the coupling. Further, we find the sign and magnitude of this shift differs for changes in electron-electron and electron-phonon coupling. This behavior should continue to hold in strongly correlated materials with narrow bandwidths.

Finally, we consider an extension of the Hatsugai-Komoto model, which has been proposed as a toy model of cuprate superconductivity. By coupling this model to a zero-momentum static phonon, we find that the solution of this non-equilibrium lattice model is identical in form to the solution of the local Holstein-Hubbard model presented here. We then explore the behavior of the first PES moment in this model in one spatial dimension, and find behavior consistent with that of the local HH model. This suggests that the first PES moment represents an experimental signature of changing subsystem couplings. In the future it may be interesting to look at simulating electric field effects in the modified HK model via a Peierls' substitution, and from this to determine a self-consistent electron-phonon coupling change. In addition it could be valuable to examine the modified HK model in the superconducting as well as the normal state.}

{In conclusion, the results presented here represent exact analytic solutions of two simple non-equilibrium electron-phonon and electron-electron coupled systems. These results have the potential to be extended to near-atomic limit calculations and strong-coupling-type calculations, but on their own they already encode interesting physical information, leading us to suggest the first PES moment as a signature of dynamic changes in subsystem couplings. }

\section*{Acknowledgements} \label{sec:acknowledgements}
This work was supported by
the Department of Energy, Office of Basic Energy Sciences, Division of Materials Sciences and Engineering under Contract No.~DE-SC0019126.
J.~K.~F. was also supported by the McDevitt bequest at Georgetown. We acknowledge useful discussions with K.~Najafi {and A.P.~Jauho}. 
\bibliography{mybib} 

\appendix 
\section{Derivation Details}
\label{sec:timedep}
In this section we provide some details of the derivation of the exact Green's functions by deriving the exact time-evolution operator for the non-equilibrium system.
\subsection{Time Evolution Operator}
To obtain the Green's functions of Eqs.~(\ref{gfdef}) and  (\ref{2partgf}), we must first determine the time evolution operator. To do this, we start by switching to the interaction representation with respect to the phonon part of the Hamiltonian and following the technique of Ref.~\onlinecite{jimHolstein}. In the interaction representation, we factorize the time evolution operator into two pieces via 
\begin{equation}
    \hat{U}(t,t_{min})=e^{-i\hat{a}^\dagger\hat{a}\omega (t-t_{min})}\mathcal{T}_t\exp\left [-i\int_{t_{min}}^t\!\!\!dt^\prime\hat{\mathcal H}_I(t^\prime)\right ]
\end{equation}
where the interaction Hamiltonian has two pieces $\hat{\mathcal H}_I=\hat{\mathcal H}_I^{el-ph}+\hat{\mathcal H}_I^{coul}$. The electron-phonon coupled piece acquires an additional time dependence through the time dependence of the phonon raising and lowering operators
\begin{widetext}
\begin{equation}
   \hat{\mathcal{ H}}^{el-ph}_I(t)=\frac{g(t)}{\sqrt{2m\omega}}\big(e^{i\omega \mathcolorbox{white}{\left(t-t_{min}\right)}} \hat{a}^{\dagger}+e^{-i\omega \mathcolorbox{white}{\left(t-t_{min} \right) }}
    \hat{a}\big)\big(\hat{n}_{\sigma}+\hat{n}_{\bsig}\big),
\end{equation}
\end{widetext}
while the Coulomb piece is unaltered ($\hat{\mathcal H}_I^{coul}(t)=U(t)\hat{n}_\uparrow\hat{n}_\downarrow$). These two pieces commute with each other, so the time evolution can be further factorized into a piece from the Coulomb part of the Hamiltonian and a piece from the electron-phonon part. The Coulomb piece is also simple, because it requires no time ordering, and is given by $\exp\left (-i\hat{n}_\uparrow\hat{n}_\downarrow\int_{t_{min}}^t\!\!\!dt^\prime U(t^\prime)\right  )$.
The electron-phonon piece is more complicated because the electron-phonon interaction Hamiltonian at two different times does not commute with itself: 
\begin{equation}
    \left [\hat{\mathcal H}^{el-ph}_I(t),\hat{\mathcal H}^{el-ph}_I(t')\right ]=\frac{g(t)g(t^\prime)}{m\omega}i\sin\omega(t-t^\prime) (\hat{n}_{\uparrow}+\hat{n}_{\downarrow})^2,
\end{equation}
but the commutator does commute with $\hat{\mathcal H}^{el-ph}_I(t^{\prime\prime})$. This allows us to solve for the time-evolution operator following the strategy used by Landau and Lifshitz\cite{landau_lifshitz} and by Gottfried\cite{gottfried} in solving the driven simple harmonic oscillator.  We first form the function 
\begin{equation}
    \hat{w}_I(t,t_{min})=\int_{t_{min}}^t\,dt'\ \hat{\mathcal H}^{el-ph}_I(t'),
\end{equation}
which satisfies the following equal-time commutator:
\begin{eqnarray}
    \left [\hat{\mathcal H}^{el-ph}_I(t),\hat{w}_I(t,t_{min})\right ]&=&i\frac{g(t)}{m\omega}\int_{t_{min}}^t\!\!\!dt^\prime g(t^\prime)\sin\omega(t-t^\prime)\nonumber\\
    &\times&(\hat{n}_{\uparrow}+\hat{n}_{\downarrow})^2.
\end{eqnarray}
This commutator commutes with all other operators in the different terms in the Hamiltonian.
When these two conditions hold, we have the following operator identity: 
\begin{align}
          &e^{i\hat{w}_I(t,t_{min})}\left(i\frac{\partial}{\partial t}-\hat{\mathcal H}^{el-ph}_{I}(t)\right )e^{-i\hat{w}_I(t,t_{min})} \nonumber\\
    &~~~~~=i\frac{\partial}{\partial t}-\frac{i}{2}\left [\hat{w}_I(t,t_{min}),\hat{\mathcal H}^{el-ph}_{I}(t)\right ].
\end{align}
This allows us to evaluate the following time-ordered product directly as 
\begin{align}
    &\mathcal{T}_t\exp\left[-i\int_{t_{min}}^t\!\!\!dt'\ \hat{\mathcal H}^{el-ph}_{I}(t')\right ]=\exp\left [-i\hat{w}_I(t,t_{min})\right ]\nonumber\\
&~~~~~\times\exp\left(\frac{1}{2}\int_{t_{min}}^t\!\!\!dt'\ \left [\hat{w}_I(t',t_{min}),\hat{\mathcal H}^{el-ph}_{I}(t')\right ]\right).
\end{align}
Applying this result to our specific Hamiltonian, yields the full time-evolution operator
\begin{widetext}
\begin{align}
     &\hat{U}(t,t_{min})=\exp\left[-i\int_{t_{min}}^t\!\!\!dt'\left (-\mu \hat{n}_f+\omega \hat{a}^{\dagger}\hat{a}+\hat{n}_{\sig}\hat{n}_{\bsig}U(t')\right )\right ]\times\exp\left[-i\int_{t_{min}}^t\!\!\!dt' \frac{g(t')}{\sqrt{2m\omega}}(e^{i\omega \left(t'\mathcolorbox{white}{-t_{min}}\right)}\hat{a}^{\dagger}+e^{-i\omega \left(t'\mathcolorbox{white}{-t_{min}}\right)}\hat{a})\hat{n}_f\right]\\ &\times\exp\left[\frac{i}{2m\omega}\int_{t_{min}}^t\!\!\!dt' \int_{t_{min}}^{t'}\!\!\!dt''\ g(t')g(t'')\sin\omega(t'-t'')\hat{n}_f^2\right]
     \nonumber
\end{align}
\end{widetext}
where $\hat{n}_f=\hat{n}_{\uparrow}+\hat{n}_{\downarrow}$ is the total electron number operator. Note that because $\hat{n}_{\sigma}^2=\hat{n}_{\sigma}$, as required by the Pauli exclusion principle, we also have that $\hat{n}_f^2=(\hat{n}_{\uparrow}+\hat{n}_{\downarrow})^2=\hat{n}_{\uparrow}+\hat{n}_{\downarrow}+2\hat{n}_\uparrow \hat{n}_\downarrow$. 
\subsection{Partition Function}
Calculating thermal averages requires us to also determine the atomic partition function, which is defined to be
\begin{align}
    \mathcal{Z}&=\text{Tr}_{b,f}\big\{\hat{U}(t=-i\beta+t_{min},t_{min})\big\}\nonumber\\&=\text{Tr}_{b,f}\big\{e^{-\beta\hat{\mathcal H}(t_{min})}\big\},
\end{align}
where the trace is over product states composed of the tensor product of the Fermionic states $\ket{0},\ket{\uparrow},\ket{\downarrow},\ket{\uparrow\downarrow}$ with the Bosonic simple harmonic oscillator number states $|n\rangle=\frac{1}{\sqrt{n!}}\left (\hat{a}^\dagger\right )^n|0\rangle$. Evaluating the trace by using standard methods, we arrive at Eq.~(\ref{partition})
with the Bose-Einstein distribution function given by Eq.~(\ref{bose}).

\subsection{Time Dependence of Field Operators}
Next we need the time dependence of the Fermionic field operators in the Heisenberg representation, 
\begin{equation}
    \hat{c}_{\sigma}(t)=\hat{U}^{\dagger}(t,t_{min})\hat{c}_{\sigma}\hat{U}(t,t_{min})
\end{equation}
and similarly for the conjugate \begin{equation}
    \hat{c}_{\sigma}^{\dagger}(t)=\hat{U}^{\dagger}(t,t_{min})\hat{c}_{\sigma}^{\dagger}(t)\hat{U}(t,t_{min}).
\end{equation}
To evaluate these operator expressions, we note that the only operator in $\hat{U}(t,t_{min})$ that does not commute with $\hat{c}_{\sigma}$ is $\hat{n}_{\sigma}$. Hence, we need to compute 
\begin{equation}
    f(\alpha)=e^{i\alpha \hat{n}_{\sigma}}\hat{c}_{\sigma}e^{-i\alpha \hat{n}_{\sigma}}.
\end{equation}
Differentiating gives us 
\begin{equation}
    \frac{df(\alpha)}{d\alpha}=e^{i\alpha \hat{n}_{\sigma}}[\hat{n}_{\sigma},\hat{c}_{\sigma}]e^{-i\alpha \hat{n}_{\sigma}} =-if(\alpha),
\end{equation}
which then can be integrated to yield
\begin{equation}
    f(\alpha)=e^{-i\alpha}\hat{c}_{\sigma}.
\end{equation}\\
Hence, we find that\\
\begin{widetext}
\begin{align}
\hat{c}_{\sigma}(t)&=\exp\left[-i\int_{t_{min}}^t\!\!\!dt'\left (-\mu +\hat{n}_{\bsig} U(t')\right )-\frac{i}{\sqrt{2m\omega}}\int_{t_{min}}^t\!\!\!dt'\ g(t')\left (e^{i\omega \left(t'\mathcolorbox{white}{-t_{min}}\right)}\hat{a}^{\dagger}+e^{-i\omega\left(t'\mathcolorbox{white}{-t_{min}}\right)}\hat{a}\right )\right]\nonumber\\&\times\exp\left[\frac{i}{2m\omega}\int_{t_{min}}^t\!\!\!dt'\ \int_{t_{min}}^{t'}\!\!\!dt''\ g(t')g(t'')\sin\omega(t'-t'')(1+2\hat{n}_{\bar{\sigma}})\right]\hat{c}_{\sigma}
\end{align}
and for the conjugate 
\begin{align}
\hat{c}_{\sigma}^{\dagger}(t)&=\exp\left[i\int_{t_{min}}^t\!\!\!dt'\ \left (-\mu+\hat{n}_{\bsig}U(t')\right ) +\frac{i}{\sqrt{2m\omega}}\int_{t_{min}}^t\!\!\!dt'\ g(t')\left (e^{i\omega \mathcolorbox{white}{\left(t'-t_{min}\right)}}\hat{a}^{\dagger}+e^{-i\omega \mathcolorbox{white}{\left(t'-t_{min}\right)}}\hat{a}\right )\right]\nonumber\\&\times\exp\left[-\frac{i}{2m\omega}\int_{t_{min}}^t\!\!\!dt'\ \int_{t_{min}}^{t'}\!\!\!dt''\ g(t')g(t'')\sin\omega(t'-t'')(1+2\hat{n}_{\bar{\sigma}})\right]\hat{c}_{\sigma}^{\dagger}.
\end{align}
\end{widetext}
With these time-dependent operators we are able to exactly evaluate the single- and two-particle Green's functions.

\section{Expectation Values}
    \label{sec:a}
    In this Appendix, we summarize the calculation of the trace needed for different time-dependent expectation values.
This includes the terms $\langle \hat{x}(t)\rangle,\langle \hat{n}_f\rangle,\langle \hat{n}_{\bsig}\rangle, \langle \hat{n}_{\sigma}\hat{n}_{\bsig}\rangle,$ $(\langle \hat{x}^2(t)\rangle-\langle \hat{x}(t)\rangle^2), (\langle \hat{x}^3(t)\rangle-\langle \hat{x}(t)\rangle ^3)$, $\langle \hat{n}_{\bsig}\hat{x}(t)\rangle$ and $\langle \hat{n}_{\bsig}\hat{x}^2(t)\rangle$. They are evaluated by using Newton's generalized binomial theorem and properties of the simple-harmonic-oscillator states, as outlined in the main paper. To begin,
(following Ref.~\onlinecite{jimHolstein}), we have:
\begin{widetext}
    \begin{equation}
    \langle \hat{x}(t)\rangle =-\langle \hat{n}_f\rangle \left (\frac{g(t_{min})}{m\omega^2}\cos\big(\omega (t-t_{min})\big)+\text{Re}\left\{ie^{-i\omega t}\int_{t_{min}}^t\!\!\!dt'\ \frac{g(t')}{m\omega}e^{i\omega t'}\right\}\right)
\end{equation} 
with \begin{equation}
\label{nf}
    \langle \hat{n}_f\rangle =\frac{2e^{\beta\mu}\exp\left[\frac{\beta g^2(t_{min})}{2m\omega^2}\right]+2e^{\beta(2\mu-U(t_{min}))}\exp\left[\frac{2\beta g^2(t_{min})}{m\omega^2}\right]}{1+2e^{\beta\mu}\exp\left[\frac{\beta g^2(t_{min})}{2m\omega^2}\right]
    +\exp\left[\beta(2\mu-U(t_{min}))\right]\exp\left[\frac{2\beta g^2(t_{min})}{m\omega^2}\right]}.
\end{equation}

Next we have
\begin{equation}
    \langle \hat{n}_{\bsig}\rangle =\frac{e^{\beta\mu}\exp\left[\frac{g^2(t_{min})\beta}{2m\omega^2}\right]+e^{\beta\left(2\mu-U\left(t_{min}\right)\right)}\exp\left[\frac{2g^2(t_{min})\beta}{m\omega^2}\right]}{1+2e^{\beta\mu}\exp\left[\frac{g^2(t_{min})\beta}{2m\omega^2}\right]
    +e^{\beta\left(2\mu-U\left(t_{min}\right)\right)}\exp \left[\frac{2g^2(t_{min})\beta}{m\omega^2}\right]}
\end{equation}
so we see, by comparison with Eq.~(\ref{nf}) that we have $\langle \hat{n}_{\bsig}\rangle=\frac{1}{2}\langle \hat{n}_{\sigma}+\hat{n}_{\bsig}\rangle=\frac{1}{2}\langle\hat{n}_f\rangle$. In addition,
\begin{align}
    \langle \hat{n}_{\sigma}\hat{n}_{\bsig}\rangle =\frac{e^{\beta\left(2\mu-U\left(t_{min}\right)\right)}\exp\left[\frac{2g^2(t_{min})\beta}{m\omega^2}\right]}{1+2e^{\beta\mu}\exp\left[\frac{g^2(t_{min})\beta}{2m\omega^2}\right]
    +\exp\left[\beta\left(2\mu-U(t_{min})\right)\right]\exp\left[\frac{2g^2(t_{min})\beta}{m\omega^2}\right]},
\end{align}
so we have \begin{align}
    \begin{split}
    \langle \hat{n}_{\bar{\sigma}}\hat{x}(t)\rangle = -\left(\frac{g(t_{min})}{m\omega^2}\cos\left(\omega\left( t-t_{min}\right)\right)+\text{Re}\left\{ie^{-i\omega t}\int_{t_{min}}^t\!dt'\ \frac{g(t')}{m\omega}e^{i\omega t'}\right\}\right)\left(\langle \hat{n}_{\sigma}\hat{n}_{\bar{\sigma}}\rangle +\langle \hat{n}_{\bar{\sigma}}\rangle\right).
\end{split}
\end{align}
Next, \begin{equation}
    \langle \hat{n}_{\bsig}\hat{x}^2(t)\rangle=\left(-\frac{g(t_{min})}{m\omega^2}\cos\left(\omega\left( t-t_{min}\right)\right)-\text{Re}\left\{ie^{-i\omega t}\int_{t_{min}}^t\!dt'\ \frac{g(t')}{m\omega}e^{i\omega t'}\right\}\right)^2\left(\langle \hat{n}_{\bsig}\rangle+3\langle \hat{n}_{\sigma}\hat{n}_{\bsig}\rangle\right)+\frac{1}{2m\omega}\coth(\frac{\beta\omega}{2})\langle \hat{n}_{\bsig}\rangle
\end{equation}

In addition, \begin{align}
    \langle \hat{x}^2(t)\rangle -\langle \hat{x}(t)\rangle^2&=
\left(\frac{g(t_{min})}{m\omega^2}\cos\left(\omega\left( t-t_{min}\right)\right)+\text{Re}\left\{ie^{-i\omega t}\int_{t_{min}}^t\!dt'\ \frac{g(t')}{m\omega}e^{i\omega t'}\right\}\right)^2\\
&\times\left(\langle \hat{n}_{\uparrow}+\hat{n}_{\downarrow}\rangle -\langle \hat{n}_{\uparrow}+\hat{n}_{\downarrow}\rangle^2 +2\langle \hat{n}_{\uparrow}\hat{n}_{\downarrow}\rangle\right)+\frac{1}{2m\omega}\coth\left(\frac{\beta\omega}{2}\right)
\end{align}
And finally,  \begin{align} \begin{split}
\langle \hat{x}^3(t)\rangle -\langle \hat{x}(t)\rangle^3 &= -\left(\frac{g(t_{min})}{m\omega^2}\cos\left(\omega\left( t-t_{min}\right)\right)+\text{Re}\left\{ie^{-i\omega t}\int_{t_{min}}^t\!dt'\ \frac{g(t')}{m\omega}e^{i\omega t'}\right\}\right)^3\left(\langle \hat{n}_{\uparrow}+\hat{n}_{\downarrow}\rangle -\langle \hat{n}_{\uparrow}+\hat{n}_{\downarrow}\rangle^3+6\langle \hat{n}_{\uparrow}\hat{n}_{\downarrow}\rangle \right)\\&+\frac{3}{2m\omega}\coth\left(\frac{\beta\omega}{2}\right) \langle \hat{x}(t)\rangle
\end{split}
\end{align}
\end{widetext}
These are all the expectation values required to evaluate the spectral moments up to third order. 

\section{Two-particle Green's function}
\label{sec:twopart}
In this Appendix, we present the exact non-equilibrium two-particle Green's function for the atomic Holstien-Hubbard model, which is calculated in the same manner as the single particle case. For the two particle Green's function, there are no known sum rules to compare our results against, so we no longer have an independent check. There are two possibilities for the spin $\sigma'$, $\sigma'=\sigma$ or $\sigma'=\bsig$. 
Here we will enumerate all the terms which appear in the two-particle Green's function, defined by
 \begin{equation}
 \label{2partgf}
    \mathcal{G}_{\sigma\sigma'}(t_0,t_1,t_2,t_3)=-\langle \mathcal{T}_t \hat{c}_{\sigma}(t_0)\hat{c}_{\sigma'}(t_1)\hat{c}_{\sigma'}^{\dagger}(t_2)\hat{c}_{\sigma}^{\dagger}(t_3)\rangle.
\end{equation}

To begin, there is a time-dependent factor which is the same for every term, so we'll define it here. 
\begin{align}
   & A(t_0,t_1,t_2,t_3)\equiv A =\frac{\big (n(\omega)+1\big )}{Z}e^{i\mu(t_0+t_1-t_2-t_3)}\nonumber\\ &\times\exp\left[\left(-\frac{1}{2}-n(\omega)\right)\left|C(t_2)+C(t_3)-C(t_0)-C(t_1)\right|^2\right]\nonumber\\ &\times\exp\big[i\big(I(t_0)+I(t_1)-I(t_2)-I(t_3)\big)\big]
\end{align}
where again $n(\omega), C(t)$ and $I(t)$ are given by Eqs. ~(\ref{bose}), ~(\ref{Coft}), and ~(\ref{Ioft}) respectively.
Next define 
\begin{align}
    F(t)\equiv2I(t)-\int_{t_{min}}^t\!dt'\ U(t')
\end{align} 
and the prefactor, which depends on the four times (and whether each time is associated with a creation or an annihilation operator),  
\begin{widetext} 
\begin{align} 
        P((\pm)_1t,(\pm)_2t',(\pm)_3\bar{t},(\pm)_4\bar{t}') & = \exp\big[-i\text{Im}\{(\pm)_1(\pm)_2C^{*}(t)C(t')
        +(\pm)_3(\pm)_4C^{*}(\bar{t})C(\bar{t}')+(\pm)_1(\pm)_3C^{*}(t)C(\bar{t})\nonumber
       \\& +(\pm)_1(\pm)_4C^{*}(t)C(\bar{t}')+(\pm)_2(\pm)_3C^{*}(t')C(\bar{t})
        +(\pm)_2(\pm)_4C^{*}(t')C(\bar{t}')\}\big],
\end{align}
\end{widetext}
where the $+$ sign is for a time associated with a creation operator, and the $-$ sign is for a time associated with an annihilation operator. Finally, we note that the sign of the term is negative (positive) if the parity of the permutation which brings the times back to the order $t_0>t_1>t_2>t_3$ is even (odd). 
\subsection{Same Spin Case}
For the case where the spins are the same, we find a number of terms in the time ordering vanish, because $\hat{c}^2_{\sigma}=\hat{c}_{\sigma}^{\dagger 2}=0$ due to the Pauli exclusion principle. In this case, only 8 of the 24 possible time-orderings are non-vanishing. These are the following eight time orderings:  \begin{enumerate}
    \item $t_0>t_2>t_1>t_3$
    \item $t_0>t_3>t_1>t_2$
    \item $t_1>t_2>t_0>t_3$
    \item $t_1>t_3>t_0>t_2$
    \item $ t_2>t_1>t_3>t_0$
    \item $t_2>t_0>t_3>t_1$
    \item $t_3>t_1>t_2>t_0$
    \item $t_3>t_0>t_2>t_1$
\end{enumerate}
When we expand the time-ordering operator for the same spin Green's function, we obtain 
\begin{widetext}
\begin{align}
\begin{split}
    \mathcal{G}_{\sigma\sigma}(t_0,t_1,t_2,t_3)&=-\langle {\mathcal T}_t\hat{c}_{\sigma}(t_0)\hat{c}_{\sigma}(t_1)\hat{c}^{\dagger}_{\sigma}(t_2)\hat{c}_{\sigma}^{\dagger}(t_3)\rangle
    \\&=\theta(t_0,t_2,t_1,t_3)\langle \hat{c}_{\sigma}(t_0)\hat{c}_{\sigma}^{\dagger}(t_2)\hat{c}_{\sigma}(t_1)\hat{c}_{\sigma}^{\dagger}(t_3)\rangle-\theta(t_0,t_3,t_1,t_2)\langle \hat{c}_{\sigma}(t_0)\hat{c}_{\sigma}^{\dagger}(t_3)\hat{c}_{\sigma}(t_1)\hat{c}_{\sigma}^{\dagger}(t_2)\rangle \nonumber\\ 
    &+\theta(t_1,t_3,t_0,t_2)\langle \hat{c}_{\sigma}(t_1)\hat{c}_{\sigma}^{\dagger}(t_3)\hat{c}_{\sigma}(t_0)\hat{c}_{\sigma}^{\dagger}(t_2)\rangle -\theta(t_1,t_2,t_0,t_3)\langle \hat{c}_{\sigma}(t_1)\hat{c}_{\sigma}^{\dagger}(t_2)\hat{c}_{\sigma}(t_0)\hat{c}_{\sigma}^{\dagger}(t_3)\rangle \nonumber\\
    &+\theta(t_2,t_0,t_3,t_1)\langle \hat{c}_{\sigma}^{\dagger}(t_2)\hat{c}_{\sigma}(t_0)\hat{c}_{\sigma}^{\dagger}(t_3)\hat{c}_{\sigma}(t_1)\rangle -\theta(t_2,t_1,t_3,t_0)\langle \hat{c}_{\sigma}^{\dagger}(t_2)\hat{c}_{\sigma}(t_1)\hat{c}_{\sigma}^{\dagger}(t_3)\hat{c}_{\sigma}(t_0)\rangle \nonumber\\
    &+\theta(t_3,t_1,t_2,t_0)\langle \hat{c}_{\sigma}^{\dagger}(t_3)\hat{c}_{\sigma}(t_1)\hat{c}_{\sigma}^{\dagger}(t_2)\hat{c}_{\sigma}(t_0)\rangle -\theta(t_3,t_0,t_2,t_1)\langle \hat{c}_{\sigma}^{\dagger}(t_3)\hat{c}_{\sigma}(t_0)\hat{c}_{\sigma}^{\dagger}(t_2)\hat{c}_{\sigma}(t_1)\rangle. \nonumber
    \end{split}
\end{align}
The expectation values are as follows:
    \begin{enumerate}
    \item $t_0>t_2>t_1>t_3$: \begin{align}
       & \langle \hat{c}_{\sigma}(t_0)\hat{c}_{\sigma}^{\dagger}(t_2)\hat{c}_{\sigma}(t_1)\hat{c}_{\sigma}^{\dagger}(t_3)\rangle = A\exp\left[P(-t_0,t_2,-t_1,t_3)\right] \bigg\{1+\exp\left[\beta\left(\mu+\frac{g^2(t_{min})}{2m\omega^2}\right)\right] \nonumber \\
       & \times\exp\left[-i\frac{2g(t_{min})}{\sqrt{2m\omega^3}}\text{Re}\left\{-C(t_0)+C(t_2)-C(t_1)+C(t_3)\right\}\right]\exp\left[-i\left(-F(t_0)+F(t_2)-F(t_1)-F(t_3)\right)\right]\bigg\}\nonumber
    \end{align}
    \item $t_0>t_3>t_1>t_2$:\begin{align}\begin{split}
         &- \langle \hat{c}_{\sigma}(t_0)\hat{c}_{\sigma}^{\dagger}(t_2)\hat{c}_{\sigma}(t_1)\hat{c}_{\sigma}^{\dagger}(t_3)\rangle = -A\exp\big[P(-t_0,t_3,-t_1,t_2)\big]\big\{1+\exp\big[\beta\big(\mu+\frac{g^2(t_{min})}{2m\omega^2}\big)\big]\\
        &\times\exp\big[-i\frac{2g(t_{min})}{\sqrt{2m\omega^3}}\text{Re}\big\{-C(t_0)+C(t_2)-C(t_1)+C(t_3)\big\}\big]\exp\big[-i\big(-F(t_0)+F(t_2)-F(t_1)-F(t_3)\big)\big]\big\}
    \end{split}
    \end{align}
    \item $t_1>t_2>t_0>t_3$:
    \begin{align}\begin{split}
        & - \langle \hat{c}_{\sigma}(t_1)\hat{c}_{\sigma}^{\dagger}(t_2)\hat{c}_{\sigma}(t_0)\hat{c}_{\sigma}^{\dagger}(t_3)\rangle = -A\exp\big[P(-t_1,t_2,-t_0,t_3)\big]\big\{1+\exp\big[\beta\big(\mu+\frac{g^2(t_{min})}{2m\omega^2}\big)\big]\\
        &\times\exp\big[-i\frac{2g(t_{min})}{\sqrt{2m\omega^3}}\text{Re}\big\{-C(t_0)+C(t_2)-C(t_1)+C(t_3)\big\}\big]\exp\big[-i\big(-F(t_0)+F(t_2)-F(t_1)-F(t_3)\big)\big]\big\}
    \end{split}
    \end{align}
    \item $t_1>t_3>t_0>t_2$: \begin{align}\begin{split}
          \langle &\hat{c}_{\sigma}(t_0)\hat{c}_{\sigma}^{\dagger}(t_2)\hat{c}_{\sigma}(t_1)\hat{c}_{\sigma}^{\dagger}(t_3)\rangle = A\exp\big[P(-t_1,t_3,-t_0,t_2)\big]\big\{1+\exp\big[\beta\big(\mu+\frac{g^2(t_{min})}{2m\omega^2}\big)\big]\\
        &\times\exp\big[-i\frac{2g(t_{min})}{\sqrt{2m\omega^3}}\text{Re}\big\{-C(t_0)+C(t_2)-C(t_1)+C(t_3)\big\}\big]\exp\big[-i\big(-F(t_0)+F(t_2)-F(t_1)-F(t_3)\big)\big]\big\}
    \end{split}
    \end{align}
    Note that for these orderings all that changed is the ordering of the times in the prefactor $P$. We see something similar for the next four orderings. 
    \item $t_2>t_1>t_3>t_0:$ \begin{align}
        \begin{split}
            &-\langle \hat{c}_{\sigma}^{\dagger}(t_2)\hat{c}_{\sigma}(t_1)\hat{c}_{\sigma}^{\dagger}(t_3)\hat{c}_{\sigma}(t_0)\rangle = -A\exp\big[P(t_2,-t_1,t_3,-t_0)\big]\bigg\{\exp\big[\beta\big(\mu+\frac{g^2(t_{min})}{2m\omega^2}\big)\big]\\
           & \times\exp\big[-i\frac{2g(t_{min})}{\sqrt{2m\omega^3}}\text{Re}\big\{-C(t_0)+C(t_2)-C(t_1)+C(t_3)\big\}\big]+\exp\big[\beta\big(2\mu-U(t_{min})+\frac{2g^2(t_{min})}{m\omega^2}\big)\big]\\
            &\times\exp\big[-i\big(-F(t_0)+F(t_2)-F(t_1)+F(t_2)\big)\big]\exp\big[-i\frac{4g(t_{min})}{\sqrt{2m\omega^3}}\text{Re}\big\{-C(t_0)+C(t_2)-C(t_1)+C(t_3)\big\}\big]\bigg\}
        \end{split}
    \end{align}
    \item $t_2>t_0>t_3>t_1$:
    \begin{align}\begin{split}
            &\langle\hat{c}_{\sigma}^{\dagger}(t_2)\hat{c}_{\sigma}(t_0)\hat{c}_{\sigma}^{\dagger}(t_3)\hat{c}_{\sigma}(t_1)\rangle = A\exp\big[P(t_2,-t_0,t_3,-t_1)\big]\bigg\{\exp\big[\beta\big(\mu+\frac{g^2(t_{min})}{2m\omega^2}\big)\big]\\
            &\times\exp\big[-i\frac{2g(t_{min})}{\sqrt{2m\omega^3}}\text{Re}\big\{-C(t_0)+C(t_2)-C(t_1)+C(t_3)\big\}\big]+\exp\big[\beta\big(2\mu-U(t_{min})+\frac{2g^2(t_{min})}{m\omega^2}\big)\big]\\
            &\times\exp\big[-i\big(-F(t_0)+F(t_2)-F(t_1)+F(t_2)\big)\big]\exp\big[-i\frac{4g(t_{min})}{\sqrt{2m\omega^3}}\text{Re}\big\{-C(t_0)+C(t_2)-C(t_1)+C(t_3)\big\}\big]\bigg\}
    \end{split}\end{align}
    \item $t_3>t_1>t_2>t_0$:\begin{align}
        \begin{split}
            &\langle \hat{c}_{\sigma}^{\dagger}(t_3)\hat{c}_{\sigma}(t_1)\hat{c}_{\sigma}^{\dagger}(t_2)\hat{c}_{\sigma}(t_0)\rangle = A\exp\big[P(t_3,-t_1,t_2,-t_0)\big]\bigg\{\exp\big[\beta\big(\mu+\frac{g^2(t_{min})}{2m\omega^2}\big)\big]\\
            &\times\exp\big[-i\frac{2g(t_{min})}{\sqrt{2m\omega^3}}\text{Re}\big\{-C(t_0)+C(t_2)-C(t_1)+C(t_3)\big\}\big]+\exp\big[\beta\big(2\mu-U(t_{min})+\frac{2g^2(t_{min})}{m\omega^2}\big)\big]\\
            &\times\exp\big[-i\big(-F(t_0)+F(t_2)-F(t_1)+F(t_2)\big)\big]\exp\big[-i\frac{4g(t_{min})}{\sqrt{2m\omega^3}}\text{Re}\big\{-C(t_0)+C(t_2)-C(t_1)+C(t_3)\big\}\big]\bigg\}
        \end{split}
    \end{align}
    \item $t_3>t_0>t_2>t_1$:\begin{align}
        \begin{split}
           &- \langle \hat{c}_{\sigma}^{\dagger}(t_3)\hat{c}_{\sigma}(t_0)\hat{c}_{\sigma}^{\dagger}(t_2)\hat{c}_{\sigma}(t_1)\rangle = -A\exp\big[P(t_3,-t_0,t_2,-t_1)\big]\bigg\{\exp\big[\beta\big(\mu+\frac{g^2(t_{min})}{2m\omega^2}\big)\big]\\
           & \times\exp\big[-i\frac{2g(t_{min})}{\sqrt{2m\omega^3}}\text{Re}\big\{-C(t_0)+C(t_2)-C(t_1)+C(t_3)\big\}\big]+\exp\big[\beta\big(2\mu-U(t_{min})+\frac{2g^2(t_{min})}{m\omega^2}\big)\big]\\
            &\times\exp\big[-i\big(-F(t_0)+F(t_2)-F(t_1)+F(t_2)\big)\big]\exp\big[-i\frac{4g(t_{min})}{\sqrt{2m\omega^3}}\text{Re}\big\{-C(t_0)+C(t_2)-C(t_1)+C(t_3)\big\}\big]\bigg\}
        \end{split}
    \end{align}
\end{enumerate}
So for this case, unlike the mixed spin case considered next, we can write a reasonably concise expression for the two particle Green's function:
\begin{align}
\begin{split}
    &\mathcal{G}_{\sigma\sigma}(t_0,t_1,t_2,t_3)=A\bigg(\bigg\{1+\exp\big[\beta\big(\mu+\frac{g^2(t_{min})}{2m\omega^2}\big)\big]\exp\big[-i\frac{2g(t_{min})}{\sqrt{2m\omega^3}}\text{Re}\big\{-C(t_0)+C(t_2)-C(t_1)+C(t_3)\big\}\big]\\&\times\exp\big[-i\big(-F(t_0)+F(t_2)-F(t_1)-F(t_3)\big)\big]\bigg\}\\
    &\times\bigg\{\theta(t_0,t_2,t_1,t_3)\exp\big[P(-t_0,t_2,-t_1,t_3)\big]-\theta(t_0,t_3,t_1,t_2)\exp\big[P(-t_0,t_3,-t_1,t_2)\big]\\&+\theta(t_1,t_3,t_0,t_2)\exp\big[P(-t_1,t_3,-t_0,t_2)\big]-\theta(t_1,t_2,t_0,t_3)\exp\big[P(-t_1,t_2,-t_0,t_3)\big]\bigg\}\\
    &+\bigg\{\exp\big[\beta\big(\mu+\frac{g^2(t_{min})}{2m\omega^2}\big)\big]\exp\big[-i\frac{2g(t_{min})}{\sqrt{2m\omega^3}}\text{Re}\big\{-C(t_0)+C(t_2)-C(t_1)+C(t_3)\big\}\big]+\exp\big[\beta\big(2\mu-U(t_{min})+\frac{2g^2(t_{min})}{m\omega^2}\big)\big]\\
            &\times\exp\big[-i\big(-F(t_0)+F(t_2)-F(t_1)+F(t_2)\big)\big]\exp\big[-i\frac{4g(t_{min})}{\sqrt{2m\omega^3}}\text{Re}\big\{-C(t_0)+C(t_2)-C(t_1)+C(t_3)\big\}\big]\bigg\}\\& \times\bigg\{\theta(t_2,t_0,t_3,t_1)\exp\big[P(t_2,-t_0,t_3,-t_1)\big]-\theta(t_2,t_1,t_3,t_0)\exp\big[P(t_2,-t_1,t_3,-t_0)\big]\\&+\theta(t_3,t_1,t_2,t_0)\exp\big[P(t_3,-t_1,t_2,-t_0)\big]-\theta(t_3,t_0,t_2,t_1)\exp\big[P(t_3,-t_0,t_2,-t_1)\big]\bigg\}\bigg)\end{split}
\end{align} 
\end{widetext}
where all the functions which appear are defined above.
\subsection{Mixed Spin Case}
For the case when $\sigma'=\bsig$, none of the orderings vanish and each of the 24 terms has a slightly different structure. We will enumerate each of the terms here. To begin, when the time-ordering operator is expanded the mixed spin Green's function is defined as\begin{widetext} \begin{align}
\begin{split}
\mathcal{G}_{\sigma\bsig}(t_0,t_1,t_2,t_3)=
-\theta(t_0,t_1,t_2,t_3)\langle \hat{c}_{\sigma}(t_0)\hat{c}_{\bsig}(t_1)c^{\dagger}_{\bsig}(t_2)\hat{c}_{\sigma}^{\dagger}(t_3)\rangle+\theta(t_1,t_0,t_2,t_3)\langle \hat{c}_{\bsig}(t_1)\hat{c}_{\sigma}(t_0)\hat{c}_{\bsig}^{\dagger}(t_2)\hat{c}_{\sigma}^{\dagger}(t_3)\rangle \\
-\theta(t_1,t_0,t_3,t_2)\langle \hat{c}_{\bsig}(t_1)\hat{c}_{\sigma}(t_0)\hat{c}_{\sigma}^{\dagger}(t_3)\hat{c}_{\bsig}^{\dagger}(t_2)\rangle +\theta(t_0,t_1,t_3,t_2)\langle \hat{c}_{\sigma}(t_0)\hat{c}_{\bsig}(t_1)c^{\dagger}_{\sigma}(t_3)c^{\dagger}_{\bsig}(t_2)\rangle \\
-\theta(t_0,t_2,t_3,t_1)\langle \hat{c}_{\sigma}(t_0)c^{\dagger}_{\bsig}(t_2)c^{\dagger}_{\sigma}(t_3)\hat{c}_{\bsig}(t_1)\rangle +\theta(t_0,t_2,t_1,t_3)\langle \hat{c}_{\sigma}(t_0)c^{\dagger}_{\bsig}(t_2)\hat{c}_{\bsig}(t_1)c^{\dagger}_{\sigma}(t_3)\rangle \\
-\theta(t_0,t_3,t_1,t_2)\langle \hat{c}_{\sigma}(t_0)c^{\dagger}_{\sigma}(t_3)\hat{c}_{\bsig}(t_1)\hat{c}_{\bsig}^{\dagger}(t_2)\rangle +\theta(t_0,t_3,t_2,t_1)\langle \hat{c}_{\sigma}(t_0)\hat{c}_{\sigma}^{\dagger}(t_3)\hat{c}_{\bsig}^{\dagger}(t_3)\hat{c}_{\bsig}(t_1)\rangle \\
-\theta(t_1,t_3,t_2,t_0)\langle \hat{c}_{\bsig}(t_1)c^{\dagger}_{\sigma}(t_3)c^{\dagger}_{\bsig}(t_2)\hat{c}_{\sigma}(t_0)\rangle +\theta(t_1,t_2,t_3,t_0)\langle \hat{c}_{\bsig}(t_1)c^{\dagger}_{\bsig}(t_2)c^{\dagger}_{\sigma}(t_3)\hat{c}_{\sigma}(t_0)\rangle \\
-\theta(t_1,t_2,t_0,t_3)\langle \hat{c}_{\bsig}(t_1)c^{\dagger}_{\bsig}(t_2)\hat{c}_{\sigma}(t_0)\hat{c}_{\sigma}^{\dagger}(t_3)\rangle +\theta(t_1,t_3,t_0,t_2)\langle \hat{c}_{\bsig}(t_1)c^{\dagger}_{\sigma}(t_3)\hat{c}_{\sigma}(t_0)c^{\dagger}_{\bsig}(t_2)\rangle \\
-\theta(t_2,t_0,t_1,t_3)\langle \hat{c}_{\bsig}^{\dagger}(t_2)\hat{c}_{\sigma}(t_0)\hat{c}_{\bsig}(t_1)\hat{c}_{\sigma}^{\dagger}(t_3)\rangle +\theta(t_2,t_1,t_0,t_3)\langle \hat{c}_{\bsig}^{\dagger}(t_2)\hat{c}_{\bsig}(t_1)\hat{c}_{\sigma}(t_0)\hat{c}_{\sigma}^{\dagger}(t_3)\rangle \\
-\theta(t_2,t_1,t_3,t_0)\langle \hat{c}_{\bsig}^{\dagger}(t_2)\hat{c}_{\bsig}(t_1)c^{\dagger}_{\sigma}(t_3)\hat{c}_{\sigma}(t_0)\rangle +\theta(t_2,t_3,t_1,t_0)\langle \hat{c}_{\bsig}^{\dagger}(t_2)c^{\dagger}_{\sigma}(t_3)\hat{c}_{\bsig}(t_1)\hat{c}_{\sigma}(t_0)\rangle \\
-\theta(t_2,t_3,t_0,t_1)\langle \hat{c}_{\bsig}^{\dagger}(t_2)c^{\dagger}_{\sigma}(t_3)\hat{c}_{\sigma}(t_0)\hat{c}_{\bsig}(t_1)\rangle +\theta(t_2,t_0,t_3,t_1)\langle \hat{c}_{\bsig}^{\dagger}(t_2)\hat{c}_{\sigma}(t_0)\hat{c}_{\sigma}^{\dagger}(t_3)\hat{c}_{\bsig}(t_1)\rangle \\
-\theta(t_3,t_1,t_0,t_2)\langle \hat{c}_{\sigma}^{\dagger}(t_3)\hat{c}_{\bsig}(t_1)\hat{c}_{\sigma}(t_0)\hat{c}_{\bsig}^{\dagger}(t_2)\rangle +\theta(t_3,t_0,t_1,t_2)\langle \hat{c}_{\sigma}^{\dagger}(t_3)\hat{c}_{\sigma}(t_0)\hat{c}_{\bsig}(t_1)\hat{c}_{\bsig}^{\dagger}(t_2)\rangle \\
-\theta(t_3,t_2,t_1,t_0)\langle \hat{c}_{\sigma}^{\dagger}(t_3),\hat{c}_{\bsig}^{\dagger}(t_2)\hat{c}_{\bsig}(t_1)\hat{c}_{\sigma}(t_0)\rangle +\theta(t_3,t_2,t_0,t_1)\langle \hat{c}_{\sigma}^{\dagger}(t_3)\hat{c}_{\bsig}^{\dagger}(t_2)\hat{c}_{\sigma}(t_0)\hat{c}_{\bsig}(t_1)\rangle \\
-\theta(t_3,t_0,t_2,t_1)\langle \hat{c}_{\sigma}^{\dagger}(t_3)\hat{c}_{\sigma}(t_0)\hat{c}_{\bsig}^{\dagger}(t_2)\hat{c}_{\bsig}(t_1)\rangle +\theta(t_3,t_1,t_2,t_0)\langle \hat{c}_{\sigma}^{\dagger}(t_3)\hat{c}_{\bsig}(t_1)\hat{c}_{\bsig}^{\dagger}(t_2)\hat{c}_{\sigma}(t_0)\rangle
\end{split}
\end{align}
\end{widetext}
In what follows, we group terms by the order of the operators $\hat{c},\hat{c}^{\dagger}$ to help with the book-keeping. There are four terms with each of the six orderings, \begin{enumerate}[label=(\roman*)]
    \item $\hat{c}\hat{c}\hat{c}^{\dagger}\hat{c}^{\dagger}$
    \item $\hat{c}\hat{c}^{\dagger}\hat{c}^{\dagger}\hat{c}$
    \item $\hat{c}\hat{c}^{\dagger}\hat{c}\hat{c}^{\dagger}$
    \item $\hat{c}^{\dagger}\hat{c}\hat{c}\hat{c}^{\dagger}$
    \item $\hat{c}^{\dagger}\hat{c}\hat{c}^{\dagger}\hat{c}$
    \item $\hat{c}^{\dagger}\hat{c}^{\dagger}\hat{c}\hat{c}$.
\end{enumerate} In evaluating these expectation values we must take more care, because the operator $\hat{n}_{\bsig}$, which appears in the time-dependence of $\hat{c}_{\sigma}(t)$ and $\hat{c}_{\sigma}^{\dagger}(t)$, no longer commutes with everything when we have operators which depend on $\bsig$. In each of the terms, only one of the four Fermionic states $\ket{0},\ket{\sigma},\ket{\bsig},\ket{\sigma\bsig}$ gives a nonzero contribution to the trace. The terms are as follows:
\begin{widetext}
\begin{enumerate}[label=(\roman*)]
    \item $\hat{c}\hat{c}\hat{c}^{\dagger}\hat{c}^{\dagger}$ terms: \\
    \begin{itemize}[label={}]
        \item 1.\underline{$t_0>t_1>t_2>t_3$}:  \begin{align}
            -\langle \hat{c}_{\sigma}(t_0)\hat{c}_{\bsig}(t_1)\hat{c}_{\bsig}^{\dagger}(t_2)\hat{c}_{\sigma}^{\dagger}(t_3)\rangle =-Ae^{P(-t_0,-t_1,t_2,t_3)} \exp\big[i\big(F(t_1)-F(t_2)\big)\big]
        \end{align}
        \item 2. \underline{$t_1>t_0>t_2>t_3$}:  \begin{align}
            \langle \hat{c}_{\bsig}(t_1)\hat{c}_{\sigma}(t_0)\hat{c}_{\bsig}^{\dagger}(t_2)\hat{c}_{\sigma}^{\dagger}(t_3)\rangle =Ae^{P(-t_1,-t_0,t_2,t_3)} \exp\big[i\big(-F(t_2)+F(t_0)\big)\big]
        \end{align}
        \item 3. \underline{$t_1>t_0>t_3>t_2$}: \begin{align}
            -\langle \hat{c}_{\bsig}(t_1)\hat{c}_{\sigma}(t_0)\hat{c}_{\sigma}^{\dagger}(t_3)\hat{c}_{\bsig}^{\dagger}(t_2)\rangle = -Ae^{P(-t_1,-t_0,t_3,t_2)} \exp[i\big(F(t_0)-F(t_3)\big)]
        \end{align}
        \item 4.\underline{$t_0>t_1>t_3>t_2$}: \begin{align}
            \langle \hat{c}_{\sigma}(t_0)\hat{c}_{\bsig}(t_1)\hat{c}_{\sigma}^{\dagger}(t_3)\hat{c}_{\bsig}^{\dagger}(t_2)\rangle =Ae^{P(-t_0,-t_1,t_3,t_2)}\exp[i\big(F(t_1)-F(t_3)\big)]
        \end{align}
    \end{itemize}
   \item $\hat{c}\hat{c}^{\dagger}\hat{c}^{\dagger}\hat{c}$: 
   \begin{itemize}[label={}]
       \item 5.\underline{$t_0>t_2>t_3>t_1$}: \begin{align}\begin{split}
           &-\langle \hat{c}_{\sigma}(t_0)\hat{c}_{\bsig}^{\dagger}(t_2)\hat{c}_{\sigma}^{\dagger}(t_3)\hat{c}_{\bsig}(t_1)=-Ae^{P(-t_0,t_2,t_3,-t_1)}\exp\big[\frac{g^2(t_{min})\beta}{2m\omega^2}\big]\\&\times\exp\big[\beta\mu-2i\frac{g(t_{min})}{\sqrt{2m\omega^3}}\text{Re}\{-C(t_0)+C(t_2)+C(t_3)-C(t_1)\}\big]\exp\big[i\big(-F(t_2)+F(t_0)\big)\big]\end{split}
       \end{align}
       \item 6.\underline{$t_0>t_3>t_2>t_1$}: \begin{align}\begin{split}
           \langle &\hat{c}_{\sigma}(t_0)\hat{c}_{\sigma}^{\dagger}(t_3)\hat{c}_{\bsig}^{\dagger}(t_2)\hat{c}_{\bsig}(t_1)\rangle =Ae^{P(-t_0,t_3,t_2,-t_1)}\exp\big[\frac{g^2(t_{min})\beta}{2m\omega^2}\big]\\&\times\exp\big[\beta\mu-2i\frac{g(t_{min})}{\sqrt{2m\omega^3}}\text{Re}\{-C(t_0)+C(t_2)+C(t_3)-C(t_1)\}\big]\exp\big[i\big(-F(t_3)+F(t_0)\big)\big]
           \end{split}
       \end{align}
       \item 7.\underline{$t_1>t_3>t_2>t_0$}: \begin{align}\begin{split}
           &-\langle \hat{c}_{\bsig}(t_1)\hat{c}_{\sigma}^{\dagger}(t_3)\hat{c}_{\bsig}^{\dagger}(t_2)\hat{c}_{\sigma}(t_0)\rangle =- Ae^{P(-t_1,t_3,t_2,-t_0)}\exp\big[\frac{g^2(t_{min})\beta}{2m\omega^2}\big]\\&\times\exp\big[\beta\mu-2i\frac{g(t_{min})}{\sqrt{2m\omega^3}}\text{Re}\{-C(t_0)+C(t_2)+C(t_3)-C(t_1)\}\big]
           \exp\big[i\big(-F(t_3)+F(t_1)\big)\big]\end{split}
       \end{align}
       \item 8. \underline{$t_1>t_2>t_3>t_0$}: \begin{align}\begin{split}
      &\langle \hat{c}_{\bsig}(t_1)\hat{c}_{\bsig}^{\dagger}(t_2)\hat{c}_{\sigma}(t_3)\hat{c}_{\sigma}(t_0)\rangle= Ae^{P(-t_1,t_3,t_2,-t_0)}\exp\big[\frac{g^2(t_{min})\beta}{2m\omega^2}\big]\\&\times\exp\big[\beta\mu-2i\frac{g(t_{min})}{\sqrt{2m\omega^3}}\text{Re}\{-C(t_0)+C(t_2)+C(t_3)-C(t_1)\}\big]\exp\big[i\big(-F(t_2)+F(t_1)\big)\big]\end{split}
       \end{align}
   \end{itemize}
   \item $\hat{c}\hat{c}^{\dagger}\hat{c}\hat{c}^{\dagger}$ terms: \\
  
   \begin{itemize}[label={}]
       \item 9.\underline{$t_0>t_2>t_1>t_3$}: \begin{align}\begin{split}
           \langle &\hat{c}_{\sigma}(t_0)\hat{c}_{\bsig}^{\dagger}(t_2)\hat{c}_{\bsig}(t_1)\hat{c}_{\sigma}^{\dagger}(t_3)\rangle = Ae^{P(-t_0,t_2,-t_1,t_3)}\exp\big[\frac{g^2(t_{min})\beta}{2m\omega^2}\big]\\&\times\exp\big[\beta\mu-2i\frac{g(t_{min})}{\sqrt{2m\omega^3}}\text{Re}\{-C(t_0)+C(t_2)+C(t_3)-C(t_1)\}\big]\exp\big[i\big(F(t_0)-F(t_2)+F(t_1)-F(t_3)\big)\big]\end{split}
       \end{align}
       \item 10. \underline{$t_0>t_3>t_1>t_2$}: \begin{align}
           -\langle \hat{c}_{\sigma}(t_0)\hat{c}_{\sigma}^{\dagger}(t_3)\hat{c}_{\bsig}(t_1)\hat{c}_{\bsig}^{\dagger}(t_2)\rangle =-Ae^{P(-t_0,t_3,-t_1,t_2)}
       \end{align}
       \item 11. \underline{$t_1>t_2>t_0>t_3$}: \begin{align}
           -\langle \hat{c}_{\bsig}(t_1)\hat{c}_{\bsig}^{\dagger}(t_2)\hat{c}_{\sigma}(t_0)\hat{c}_{\sigma}^{\dagger}(t_3)\rangle =-Ae^{P(-t_1,t_2,-t_0,t_3)}
       \end{align}
       \item 12. \underline{$t_1>t_3>t_0>t_2$}: \begin{align}\begin{split}
           &\langle \hat{c}_{\bsig}(t_1)\hat{c}_{\sigma}^{\dagger}(t_3)\hat{c}_{\sigma}(t_0)\hat{c}_{\bsig}^{\dagger}(t_2)\rangle = Ae^{P(-t_1,t_3,-t_0,t_2)}\exp\big[\frac{g^2(t_{min})\beta}{2m\omega^2}\big]\\&\times\exp\big[\beta\mu-2i\frac{g(t_{min})}{\sqrt{2m\omega^3}}\text{Re}\{-C(t_0)+C(t_2)+C(t_3)-C(t_1)\}\big] \exp\big[i\big(F(t_1)-F(t_3)+F(t_0)-F(t_2)\big)\big]\end{split}
       \end{align}
   \end{itemize}
   \item $\hat{c}^{\dagger}\hat{c}\hat{c}\hat{c}^{\dagger}$ terms: \\
   \begin{itemize}[label={}]
       \item 13.\underline{$t_2>t_0>t_1>t_3$}: \begin{align}\begin{split}
           &-\langle \hat{c}_{\bsig}^{\dagger}(t_2)\hat{c}_{\sigma}(t_0)\hat{c}_{\bsig}(t_1)\hat{c}_{\sigma}^{\dagger}(t_3)\rangle =-Ae^{P(t_2,-t_0,-t_1,t_3)}\exp\big[\frac{g^2(t_{min})\beta}{2m\omega^2}\big]\\&\times\exp\big[\beta\mu-2i\frac{g(t_{min})}{\sqrt{2m\omega^3}}\text{Re}\{-C(t_0)+C(t_2)+C(t_3)-C(t_1)\}\big]\exp[i\big(F(t_1)-F(t_3)\big)]\end{split}
       \end{align}
       \item 14. \underline{$t_2>t_1>t_0>t_3$}: \begin{align}\begin{split}
          &\langle \hat{c}_{\bsig}^{\dagger}(t_2)\hat{c}_{\bsig}(t_1)\hat{c}_{\sigma}(t_0)\hat{c}_{\sigma}^{\dagger}(t_3)\rangle= Ae^{P(t_2,-t_1,-t_0,t_3)}\exp\big[\frac{g^2(t_{min})\beta}{2m\omega^2}\big]\\&\times\exp\big[\beta\mu-2i\frac{g(t_{min})}{\sqrt{2m\omega^3}}\text{Re}\{-C(t_0)+C(t_2)+C(t_3)-C(t_1)\}\big]\exp[i\big(F(t_0)-F(t_3)\big)]\end{split}
       \end{align}
       \item 15. \underline{$t_3>t_1>t_0>t_2$}: \begin{align}\begin{split}
           &-\langle \hat{c}_{\sigma}^{\dagger}(t_3)\hat{c}_{\bsig}(t_1)\hat{c}_{\sigma}(t_0)\hat{c}_{\bsig}^{\dagger}(t_2)\rangle =-Ae^{P(t_3,-t_1,-t_0,t_2)}\exp\big[\frac{g^2(t_{min})\beta}{2m\omega^2}\big]\\&\times\exp\big[\beta\mu-2i\frac{g(t_{min})}{\sqrt{2m\omega^3}}\text{Re}\{-C(t_0)+C(t_2)+C(t_3)-C(t_1)\}\big] \exp[i\big(F(t_0)-F(t_2)\big)]\end{split}
       \end{align}
       \item 16.\underline{$t_3>t_0>t_1>t_2$} \begin{align}\begin{split}
           &\langle \hat{c}_{\sigma}^{\dagger}(t_3)\hat{c}_{\sigma}(t_0)\hat{c}_{\bsig}(t_1)\hat{c}_{\bsig}^{\dagger}(t_2)\rangle = Ae^{P(t_3,-t_0,-t_1,t_2)}\exp\big[\frac{g^2(t_{min})\beta}{2m\omega^2}\big]\\&\times\exp\big[\beta\mu-2i\frac{g(t_{min})}{\sqrt{2m\omega^3}}\text{Re}\{-C(t_0)+C(t_2)+C(t_3)-C(t_1)\}\big]\exp[i\big(F(t_1)-F(t_2)\big)]\end{split}
       \end{align}
   \end{itemize}
   \item $\hat{c}^{\dagger}\hat{c}\hat{c}^{\dagger}\hat{c}$ terms. \\
 \begin{itemize}[label={}]
       \item 17. \underline{$t_2>t_1>t_3>t_0$}: \begin{align}\begin{split}
           &-\langle \hat{c}_{\bsig}^{\dagger}(t_2)\hat{c}_{\bsig}(t_1)\hat{c}_{\sigma}^{\dagger}(t_3)\hat{c}_{\sigma}(t_0)\rangle =-Ae^{P(t_3,-t_0,t_2,-t_1)}\exp\big[\frac{2g^2(t_{min})\beta}{m\omega^2}\big]\exp\big[-\beta U(t_{min})\big]\\&\times\exp\big[2(\beta\mu-2i\frac{g(t_{min})}{\sqrt{2m\omega^3}}\text{Re}\{-C(t_0)+C(t_2)+C(t_3)-C(t_1)\})\big] \exp[i\big(-F(t_2)+F(t_1)-F(t_3)+F(t_0)\big)]\end{split}
       \end{align}
       \item 18. \underline{$t_2>t_0>t_3>t_1$}: \begin{align}\begin{split}
           &\langle \hat{c}_{\bsig}^{\dagger}(t_2)\hat{c}_{\sigma}(t_0)\hat{c}_{\sigma}^{\dagger}(t_3)\hat{c}_{\bsig}(t_1)\rangle =Ae^{P(t_2,-t_0,t_3,-t_1)}\exp\big[\frac{g^2(t_{min})\beta}{2m\omega^2}\big]\exp\big[\beta\mu-2i\frac{g(t_{min})}{\sqrt{2m\omega^3}}\text{Re}\{-C(t_0)+C(t_2)+C(t_3)-C(t_1)\}\big]\end{split}
       \end{align}
       \item 19. \underline{$t_3>t_0>t_2>t_1$}: \begin{align}\begin{split}
           &-\langle \hat{c}_{\sigma}^{\dagger}(t_3)\hat{c}_{\sigma}(t_0)\hat{c}_{\bsig}^{\dagger}(t_2)\hat{c}_{\bsig}(t_1)\rangle =-Ae^{P(t_3,-t_0,t_2,-t_1)}\exp\big[\frac{2g^2(t_{min})\beta}{m\omega^2}\big]\exp\big[-\beta U(t_{min})\big]\\& \times\exp\big[2(\beta\mu-2i\frac{g(t_{min})}{\sqrt{2m\omega^3}}\text{Re}\{-C(t_0)+C(t_2)+C(t_3)-C(t_1)\})\big]\exp[i\big(-F(t_3)+F(t_0)-F(t_2)+F(t_1)\big)]\end{split}
       \end{align}
       \item 20. \underline{$t_3>t_1>t_2>t_0$}: \begin{align}\begin{split}
           \langle \hat{c}_{\sigma}^{\dagger}(t_3)\hat{c}_{\bsig}(t_1)\hat{c}_{\bsig}^{\dagger}(t_2)\hat{c}_{\sigma}(t_0)\rangle =Ae^{P(t_3,-t_1,t_2,-t_0)}\exp\big[\frac{g^2(t_{min})\beta}{2m\omega^2}\big]\exp\big[\beta\mu-2i\frac{g(t_{min})}{\sqrt{2m\omega^3}}\text{Re}\{-C(t_0)+C(t_2)+C(t_3)-C(t_1)\}\big]\end{split}
       \end{align}
   \end{itemize}
   \item $\hat{c}^{\dagger}\hat{c}^{\dagger}\hat{c}\hat{c}$ terms: \\
  \begin{itemize}[label={}]
       \item 21. \underline{$t_2>t_3>t_0>t_1$}: \begin{align}\begin{split}
       &-\langle \hat{c}_{\bsig}^{\dagger}(t_2)\hat{c}_{\sigma}^{\dagger}(t_3)\hat{c}_{\sigma}(t_0)\hat{c}_{\bsig}(t_1)\rangle =-Ae^{P(t_2,t_3,-t_0,-t_1)}\exp\big[\frac{2g^2(t_{min})\beta}{m\omega^2}\big]\exp\big[-\beta U(t_{min})\big]\\& \times\exp\big[2(\beta\mu-2i\frac{g(t_{min})}{\sqrt{2m\omega^3}}\text{Re}\{-C(t_0)+C(t_2)+C(t_3)-C(t_1)\})\big]\exp[i\big(F(t_1)-F(t_2)\big)]\end{split}
       \end{align}
       \item 22. \underline{$t_2>t_3>t_1>t_0$}: \begin{align}\begin{split}
          &\langle \hat{c}_{\bsig}^{\dagger}(t_2)\hat{c}_{\sigma}^{\dagger}(t_3)\hat{c}_{\bsig}(t_1)\hat{c}_{\sigma}(t_1)\rangle =Ae^{P(t_2,t_3,-t_1,-t_0)}\exp\big[\frac{2g^2(t_{min})\beta}{m\omega^2}\big]\exp\big[-\beta U(t_{min})\big]\\&\times\exp\big[2(\beta\mu-2i\frac{g(t_{min})}{\sqrt{2m\omega^3}}\text{Re}\{-C(t_0)+C(t_2)+C(t_3)-C(t_1)\})\big] \exp[i\big(F(t_1)-F(t_2)\big)]\end{split}
       \end{align}
       \item 23. \underline{$t_3>t_2>t_1>t_0$}: \begin{align}\begin{split}
           &-\langle \hat{c}_{\sigma}^{\dagger}(t_3)\hat{c}_{\bsig}^{\dagger}(t_2)\hat{c}_{\bsig}(t_1)\hat{c}_{\sigma}(t_0)\rangle =-Ae^{P(t_3,t_2,-t_1,-t_0)}\exp\big[-\beta U(t_{min})\big]\exp\big[\frac{2g^2(t_{min})\beta}{m\omega^2}\big]\\&\times\exp\big[2(\beta\mu-2i\frac{g(t_{min})}{\sqrt{2m\omega^3}}\text{Re}\{-C(t_0)+C(t_2)+C(t_3)-C(t_1)\})\big] \exp[i\big(F(t_0)-F(t_3)\big)]\end{split}
       \end{align}
       \item 24. \underline{$t_3>t_2>t_0>t_1$}:
       \begin{align}\begin{split}
           &\langle \hat{c}_{\sigma}^{\dagger}(t_3)\hat{c}_{\bsig}(t_2)\hat{c}_{\sigma}(t_0)\hat{c}_{\bsig}(t_1)\rangle =Ae^{P(t_3,t_2,-t_0,-t_1)}\exp\big[\frac{2g^2(t_{min})\beta}{m\omega^2}\big]\exp\big[-\beta U(t_{min})\big]\\&\times\exp\big[2(\beta\mu-2i\frac{g(t_{min})}{\sqrt{2m\omega^3}}\text{Re}\{-C(t_0)+C(t_2)+C(t_3)-C(t_1)\})\big]\exp\big[i\big(F(t_1)-F(t_3)\big)]\end{split}
       \end{align}
   \end{itemize}
\end{enumerate}
\end{widetext}
This fully determines the two-particle Green's function in the mixed spin case. This, combined with the expression found for when the spins are the same, completely determines the exact two-particle time ordered Green's function of the local Holstein-Hubbard model with arbitrary time-dependent couplings.

\end{document}

\title{Supplemental Material for ``Exact solution of two simple non-equilibrium electron-phonon and electron-electron coupled systems: the atomic limit of the Holstein-Hubbard model and the generalized Hatsugai-Komoto model"}

\author{R.~D.~Nesselrodt and J.~K.~Freericks}
    \email[Correspondence email address: ]{rdn11@georgetown.edu}% Your name
    \affiliation{Department of Physics, Georgetown University,
              37th and O Sts. NW, Washington, D.C. 20057 USA
              %Tel.: 202-687-5984\\
              }

\date{\today} % Leave empty to omit a date

\maketitle

\section{Analytic Holstein-Hubbard Green's function Derivatives}
In this section, we analytically confirm the first four moment sum rules of the non-equilibrium Holstein-Hubbard Green's function by directly taking derivatives of the Green's function (Eq. (17) from the main text) and comparing to the expressions for the moments with $t_{ij}=0$ from Ref. 34. Our strategy is as follows: first, we will give the definitions relevant to the derivation. Second, we define general functions which bundle the time-dependence of the problem. Next we will implicitly differentiate these functions to have the final form the derivatives take, and take the limit $t_r\to 0$ of this expression. Then we explicitly take the derivatives of these functions we've defined, take the limit $t_r\to 0$, plug these back into the general form derivatives, and evaluate the moments.
\subsection{Background Definitions}
To begin, we redefine the two functions which appear frequently in our expressions, \begin{equation}
    I(t)=\frac{1}{2m\omega}\int_{t_{min}}^t\,dt'\ \int_{t_{min}}^{t'}\,dt''\ g(t')g(t'') \sin\big(\omega(t'-t'')\big)
\end{equation}
\begin{equation}
    C(t)=\frac{1}{\sqrt{2m\omega}}\int_{t_{min}}^t\,dt'\ g(t')e^{i\omega (t'-t_{min})}
\end{equation}

It's much easier to compare the derivatives to the expressions given for the moments by taking the derivative after we've taken the bosonic trace, but before we've taken the fermionic trace. At this stage, the Green's function we've derived is given by \begin{align}
\label{gresult}
    &g_{\sigma}^R(t_1,t_2)=-i\theta(t_1-t_2)\frac{\left(n(\omega)+1\right)}{\mathcal{Z}}e^{i\mu(t_1-t_2)}\exp\left[-\left(\frac{1}{2}+n(\omega)\right)\left|C(t_1)-C(t_2)\right|^2\right] \nonumber\\
    &\times\text{Tr}_f\bigg\{\exp\left[\frac{g^2(t_{min})\beta}{2m\omega^2}\hat{n}_f^2+\beta\mu\hat{n}_f-\beta U(t_{min})\hat{n}_{\sig}\hat{n}_{\bsig}\right] \exp\left[i\left(I(t_1)-I(t_2)\right)\left(1+2\hat{n}_{\bar{\sigma}}\right)\right]\exp\left[\frac{2ig(t_{min})}{\sqrt{2m\omega^3}}\hat{n}_f\text{Re}\{C(t_1)-C(t_2)\}\right]\nonumber\\
    &\times\exp\left[-i\int_{t_2}^{t_1}\,dt\ U(t)\hat{n}_{\bsig}\right]\left(e^{i\text{Im}\{C^{*}(t_1)C(t_2)\}}\hat{c}_{\sigma}\hat{c}_{\sigma}^{\dagger}+e^{-i\text{Im}\{C^{*}(t_1)C(t_2)\}}\hat{c}_{\sigma}^{\dagger}\hat{c}_{\sigma}\right)\bigg\}
\end{align}
where \begin{align}
    n(\omega)=\frac{1}{e^{\beta\omega}-1}.
\end{align}
Finally we need the expectation value of the position operator, which is\begin{align}
    \langle \hat{x}(t)\rangle =-\bigg(\frac{g(t_{min})}{m\omega^2}\cos\left(\omega\left( t-t_{min}\right)\right)+\text{Re}\big\{ie^{-i\omega t}\int_{t_{min}}^t\,dt'\ \frac{g(t')}{m\omega}e^{i\omega t'}\big\}\bigg)\langle \hat{n}_{\sig}+\hat{n}_{\bsig}\rangle.
\end{align}
The spectral moments of the Green's function are defined as \begin{align*}
    \mu^{Rn}_{at,\sigma}(t_1,t_2) = -\text{Im}\big\{i^n\frac{\partial^n}{\partial t_r^n}g^R_{at,\sigma}(t_1,t_2)\big\}\vert_{t_r\to 0}
\end{align*}
where $t_1=t_a+\frac{t_r}{2}, t_2=t_a-\frac{t_r}{2}$. 
\subsection{Derivatives Derivation}
\subsubsection{General Form Definitions}
To begin, note that everything in the exponents commutes with everything else, so we can combine and separate as needed. To begin, we'll calculate the derivatives of a general expression which has the form of the Green's function we have. Then we'll look at the derivatives of the individual terms, and finally put it all together. To begin, returning to our expression for $g^R$,writing it a little bit differently, recalling $t_1=t_a+t_r/2, t_2=t_a-t_r/2$, we have \begin{align}
     &g_{\sigma}^R(t_1,t_2)=-i\theta(t_1-t_2)\frac{\left(n(\omega)+1\right)}{\mathcal{Z}}\text{Tr}_f\bigg\{\exp\left[\frac{g^2(t_{min})\beta}{2m\omega^2}\hat{n}_f^2+\beta\mu\hat{n}_f-\beta U(t_{min})\hat{n}_{\sig}\hat{n}_{\bsig}\right]\nonumber\\
     &\exp\bigg[i\mu(t_1-t_2)-\big(\frac{1}{2}+n(\omega)\big)\big|C(t_1)-C(t_2)\big|^2
     +i\left(I(t_1)-I(t_2)\right)\left(1+2\hat{n}_{\bar{\sigma}}\right)+\frac{2ig(t_{min})}{\sqrt{2m\omega^3}}\hat{n}_f\text{Re}\{C(t_1)-C(t_2)\}\nonumber\\&-i\int_{t_1}^{t_2}\,dt\ U(t)\hat{n}_{\bsig}\bigg]
     \times\bigg(e^{i\text{Im}\{C^{*}(t_1)C(t_2)\}}\hat{c}_{\sigma}\hat{c}_{\sigma}^{\dagger}+e^{-i\text{Im}\{C^{*}(t_1)C(t_2)\}}\hat{c}_{\sigma}^{\dagger}\hat{c}_{\sigma}\bigg)\bigg\}\nonumber\\
     &\equiv -i\theta(t_1-t_2)\frac{(n(\omega)+1)}{\mathcal{Z}}\text{Tr}_f\bigg\{\exp\big[\frac{g^2(t_{min})\beta}{2m\omega^2}\hat{n}_f^2+\beta\mu\hat{n}_f-\beta U(t_{min})\hat{n}_{\bsig}\big]\exp\big[\hat{f}(t_1,t_2)]\big(e^{u(t_1,t_2)}\hat{c}_{\sigma}\hat{c}_{\sigma}^{\dagger}+e^{-u(t_1,t_2)}\hat{c}_{\sigma}^{\dagger}\hat{c}_{\sigma}\big)\bigg\}\nonumber\\
     &\equiv -i\theta(t_1-t_2)\frac{(n(\omega)+1)}{\mathcal{Z}}\text{Tr}_f\bigg\{\exp\big[\frac{g^2(t_{min})\beta}{2m\omega^2}\hat{n}_f^2+\beta\mu\hat{n}_f-\beta U(t_{min})\hat{n}_{\bsig}\big]\hat{F}(t_1,t_2)\bigg\}
\end{align}
where we've defined functions which bundle all the time-dependence of the problem,\begin{align}
    &\hat{f}(t_a+t_r/2,t_a-t_r/2)\equiv i\mu t_r-\big(\frac{1}{2}+n(\omega)\big)\big|C(t_a+t_r/2)-C(t_a-t_r/2)\big|^2-i\int_{t_2}^{t_1}\,dt\ U(t)\hat{n}_{\bsig}\nonumber\\
    &+i\left(I(t_a+t_r/2)-I(t_a-t_r/2)\right)(1+2\hat{n}_{\bsig})+\frac{2ig(t_{min})}{\sqrt{2m\omega^3}}\hat{n}_f\text{Re}\{C(t_a+t_r/2)-C(t_a-t_r/2)\}
\end{align}
and \begin{align}
    u(t_a+t_r/2,t_a-t_r/2)=i\text{Im}\{C^{*}(t_a+t_r/2)C(t_a-t_r/2)\}
\end{align} so that we can take derivatives of these. Next for notational convenience we'll drop the hat on these functions and write these functions as a function of only one time, with the understanding these depend on both average and relative times but we are only taking the derivative with respect to the relative time, i.e. \begin{align}
    \hat{f}(t_a+t_r/2,t_a-t_r/2)=f(t_r),\\ u(t_a+t_r/2,t_a-t_r/2)=u(t_r),\\ \hat{F}(t_a+t_r/2,t_a-t_r/2)=F(t_r).
    \end{align}

\subsubsection{General Form Implicit Derivatives}
We've made definitions of functions which hold all the time dependence of our Green's function. Now we take the derivatives of this general function using the chain rule. We'll use primes to denote derivatives with respect to relative time. We have the function \begin{align}
    F(t_r)=e^{f(t_r)}(e^{u(t_r)}cc^{\dagger}+e^{-u(t_r)}c^{\dagger}c)
\end{align}
which we wish to differentiate thrice. For the first derivative we have
 \begin{align}
        F'(t_r)=f'e^f(e^ucc^{\dagger}+e^{-u}c^{\dagger}c)+e^f(u'e^ucc^{\dagger}-u'e^{-u}c^{\dagger}c).
    \end{align}
 For the second derivative, we have \begin{align}
        &F''(t_r)=(f''+f'^2)e^f(e^ucc^{\dagger}+e^{-u}c^{\dagger}c)+2f'e^f(u'e^ucc^{\dagger}-u'e^{-u}c^{\dagger}c)\nonumber\\ &+e^f([u''+u'^2]e^ucc^{\dagger}+[u'^2-u'']e^{-u}c^{\dagger}c).
    \end{align}
  And finally, for the third derivative we obtain \begin{align}
        &F'''(t_r)=(f'''+3f'f''+f'^3)e^f(e^ucc^{\dagger}+e^{-u}c^{\dagger}c)+3(f''+f'^2)e^f(u'e^ucc^{\dagger}-u'e^{-u}c^{\dagger}c)\nonumber\\
        &+3f'e^f([u''+u'^2]e^ucc^{\dagger}+[u'^2-u'']e^{-u}c^{\dagger}c)+e^f([u'''+3u'u''+u'^3]e^ucc^{\dagger}-[u'^3-3u''u'+u''']e^{-u}c^{\dagger}c).
    \end{align}

\subsubsection{Limit $t_r\to0$ - Implicit Derivatives}
Note, in the limit $t_r\to 0$, which we take after taking the derivatives, both $f(t_a)=u(t_a)=0$, so $e^u=e^{-u}=e^f=1$. This is obvious for $f$. For $u$ this occurs because we have something times its complex conjugate, which must be real, then we take the imaginary part. Next note \begin{align}
    \{\hat{c}_{\sigma},\hat{c}_{\sigma}^{\dagger}\}_+= 1, \\
    \hat{c}_{\sigma}\hat{c}_{\sigma}^{\dagger}-\hat{c}_{\sigma}^{\dagger}\hat{c}_{\sigma} = 1-2\hat{n}_{\sigma}
\end{align}
With these we can take the $t_r=0$ limit of each of the derivatives above, where the primes have an implied limit $t_r\to 0$. For the first derivative we have \begin{align}\label{first}
    F'(t_r\to 0) =f' +u'(1-2\hat{n}_{\sigma}).
    \end{align}
  Next we obtain \begin{align}
    \label{second}
     F''(t_r\to 0)= f''+f'^2+2f'u'(1-2\hat{n}_{\sigma})+u'^2+u''(1-2\hat{n}_{\sigma}).
    \end{align}
  Finally for the third derivative in the relative time to zero limit, we have \begin{align}
    \label{third}
        &F'''(t_r\to 0)=(f'''+3f'f''+f'^3)+3(f''+f'^2)u'(1-2\hat{n}_{\sigma})\nonumber\\
        &+3f'u'^2+3f'u''(1-2\hat{n}_{\sigma})+3u'u''+(u'^3+u''')(1-2\hat{n}_{\sigma}).
    \end{align}

The remainder of the derivation is determining the derivatives $f',f'',f'''$ and $u', u'', u'''$ evaluated at the $t_r$ to zero limit. Then we plug into the expressions above, and evaluate the trace.
\subsection{Derivatives of the sub-functions, $f$ and $u$}

\subsubsection{\textbf{Derivatives of $u$}}
We wish to differentiate $u(t_1,t_2)=i\text{Im}\{C^{*}(t_1)C(t_2)\}$, where $t_1=t_a+\frac{t_r}{2}, t_2=t_a-\frac{t_r}{2}$. To begin we have
\begin{align}
    \frac{d C^{*}(t_1)}{dt_r}=\frac{1}{\sqrt{2m\omega}}\frac{d}{dt_r}\int_{t_{min}}^{t_a+t_r/2}\,dt\ g(t)e^{-i\omega( t-t_{min})}=\frac{1}{2\sqrt{2m\omega}}g(t_a+t_r/2)e^{-i\omega(t_a+t_r/2-t_{min})}
\end{align} and \begin{align}
    \frac{dC(t_2)}{dt_r}=\frac{1}{\sqrt{2m\omega}}\frac{d}{dt_r}\int_{t_{min}}^{t_a-t_r/2}\,dt\ g(t)e^{i\omega (t-t_{min})}=-\frac{1}{2\sqrt{2m\omega}}g(t_a-t_r/2)e^{i\omega(t_a-t_r/2-t_{min})}.
\end{align} We can now differentiate the products. For the first derivative we have \begin{align}&\frac{d}{dt_r}\big(C^{*}(t_1)C(t_2)\big)=\frac{1}{4m\omega}\bigg\{g(t_a+t_r/2)e^{-i\omega(t_a+t_r/2-t_{min})}\int_{t_{min}}^{t_a-t_r/2}\,dt\ g(t)e^{i\omega (t-t_{min})}\nonumber\\
&-g(t_a-t_r/2)e^{i\omega(t_a-t_r/2-t_{min})}\int_{t_{min}}^{t_a+t_r/2}\,dt\ g(t)e^{-i\omega (t-t_{min})}\bigg\}.
\end{align}
Taking the limit $t_r$ to zero we have
    \begin{align}
       & u'(t_r\to 0)= i\text{Im}\big\{\frac{d}{dt_r}\big(C^{*}(t_1)C(t_2)\big)\big|_{t_r\to 0}\big\} \nonumber\\
       & =\frac{i}{2m\omega}g(t_a)\int_{t_{min}}^{t_a}\sin\big(\omega(t-t_a)\big) ,   \end{align}
        or equivalently (rewriting the sine function) \begin{align}
         u'(t_r\to 0)=-\frac{i}{2m\omega}g(t_a)\text{Re}\bigg\{ie^{-i\omega t_a}\int_{t_{min}}^{t_a}\,dt\ g(t)e^{i\omega t}\bigg\}.       \end{align}

 For the second derivative we have \begin{align}
    &\frac{d^2}{dt_r^2}\big(C^{*}(t_1)C(t_2)\big)=\frac{1}{8m\omega}\bigg\{\frac{dg(t_a+t_r/2)}{dt_a}e^{-i\omega(t_a+t_r/2-t_{min})}\int_{t_{min}}^{t_a-t_r/2}\,dt\ g(t)e^{i\omega (t-t_{min})}\nonumber\\
    &+\frac{dg(t_a-t_r/2)}{dt_a}e^{i\omega(t_a-t_r/2-t_{min})}\int_{t_{min}}^{t_a+t_r/2}\,dt\ g(t)e^{-i\omega (t-t_{min})}-i\omega g(t_a+t_r/2)e^{-i\omega(t_a+t_r/2-t_{min})}\int_{t_{min}}^{t_a-t_r/2}\,dt\ g(t)e^{i\omega (t-t_{min})}\nonumber\\&+i\omega g(t_a-t_r/2)e^{i\omega(t_a-t_r/2-t_{min})}\int_{t_{min}}^{t_a+t_r/2}\,dt\ g(t)e^{-i\omega (t-t_{min})}-2g(t_a+t_r/2)g(t_a-t_r/2)e^{-i\omega t_r}\bigg\}.
\end{align}
Note that the derivatives of $g(t)$ become derivatives of $t_a$ once the limit $t_r\to 0$ is taken. So in the limit $t_r$ to zero we have  \begin{align}
&u''(t_r\to 0) = i \text{Im}\bigg\{\frac{1}{4m\omega}\big(\frac{dg(t_a)}{dt_a}\int_{t_{min}}^{t_a}\,dt\ g(t)\cos\big(\omega(t-t_a)\big)-g^2(t_a)\big)+\frac{1}{4m}g(t_a)\int_{t_{min}}^{t_a}\,dt\ g(t)\sin(\omega(t-t_a))\bigg\}.
    \end{align}
    Note this is taking the imaginary part of something real, so we have proved \begin{align}
        u''(t_r\to 0)=0.
    \end{align}
For the third derivative we have \begin{align}
    &\frac{d^3}{dt_r^3}\big(C^{*}(t_1)C(t_2)\big)=\frac{1}{8m\omega}\bigg\{\frac{1}{2}\frac{d^2g(t_a+t_r/2)}{dt_a^2}e^{-i\omega(t_a+t_r/2-t_{min})}\int_{t_{min}}^{t_a-t_r/2}\,dt\ g(t)e^{i\omega (t-t_{min})}\nonumber\\
    &-\frac{1}{2}\frac{d^2g(t_a-t_r/2)}{dt_a^2}e^{i\omega(t_a-t_r/2-t_{min})}\int_{t_{min}}^{t_a+t_r/2}\,dt\ g(t) e^{-i\omega (t-t_{min})}\nonumber\\ &-i\omega\frac{dg(t_a+t_r/2)}{dt_a}e^{-i\omega(t_a+t_r/2-t_{min})}\int_{t_{min}}^{t_a-t_r/2}\,dt\ g(t)e^{i\omega (t-t_{min})}\nonumber\\
    &-i\omega \frac{dg(t_a-t_r/2)}{dt_a}e^{i\omega(t_a-t_r/2-t_{min})}\int_{t_{min}}^{t_a+t_r/2}\,dt\ g(t) e^{i\omega (t-t_{min})}-\frac{1}{2}\frac{dg(t_a+t_r/2)}{dt_a}g(t_a-t_r/2)e^{-i\omega t_r}\nonumber\\
    &+\frac{1}{2}\frac{dg(t_a-t_r/2)}{dt_a}g(t_a+t_r/2)e^{-i\omega t_r}-\frac{\omega^2}{2}g(t_a+t_r/2)e^{-i\omega(t_a+t_r/2-t_{min})}\int_{t_{min}}^{t_a-t_r/2}\,dt\ g(t)e^{i\omega (t-t_{min})}\nonumber\\
    &+\frac{\omega^2}{2}g(t_a-t_r/2)e^{i\omega(t_a-t_r/2-t_{min})}\int_{t_{min}}^{t_a+t_r/2}\,dt\ g(t)e^{-i\omega (t-t_{min})}+3i\omega g(t_a+t_r/2)g(t_a-t_r/2)e^{-i\omega t_r}\nonumber\\
    &-\frac{dg(t_a+t_r/2)}{dt_a}g(t_a-t_r/2)e^{-i\omega t_r}+g(t_a+t_r/2)\frac{dg(t_a-t_r/2)}{dt_a}e^{-i\omega t_r}.
\end{align}
When we take the limit $t_r\to 0$ we have \begin{align}
        &u'''(t_r\to 0)= \frac{i}{8m\omega}\bigg\{\frac{d^2g(t_a)}{dt_a^2}\int_{t_{min}}^{t_a}\,dt\ g(t) \sin\big(\omega(t-t_a)\big)\nonumber\\
        &-2\omega\frac{dg(t_a)}{dt_a}\int_{t_{min}}^{t_a}\,dt\ g(t)\cos\big(\omega(t-t_a)\big)\nonumber\\
        &-\omega^2g(t_a)\int_{t_{min}}^{t_a}\,dt\ g(t)\sin\big(\omega(t-t_a)\big)+3\omega g^2(t_a)\bigg\}.
    \end{align}
So we have now determined all the derivatives $u',u'',u'''$ in the $t_r\to 0$ limit.
\subsubsection{\textbf{Derivatives of $f$}}
Next we take the derivatives of $f$, and the limit $t_r$ to zero. These are sufficiently more cumbersome, so we look at the pieces individually, then put the results together.
 \begin{enumerate}
\item \textbf{First derivatives}:\\ We have for the first non-trivial term of $f$ \begin{align}
    &\frac{d}{dt_r}\big(-\frac{1}{2}-n(\omega)\big)\big|C(t_a+t_r/2)-C(t_a-t_r/2)\big|^2=\nonumber\\&\big(-\frac{1}{2}-n(\omega)\big)\frac{1}{2}\bigg[g(t_a+t_r/2)\big(e^{-i\omega(t_a+t_r/2-t_{min})}\int_{t_a-t_r/2}^{t_a+t_r/2}\,dt\ g(t) e^{i\omega (t-t_{min})}+e^{i\omega(t_a+t_r/2-t_{min})}\int_{t_a-t_r/2}^{t_a+t_r/2}\,dt\ g(t)e^{-i\omega (t-t_{min})}\big)
    \nonumber\\
    &-g(t_a-t_r/2-t_{min})\big(e^{-i\omega(t_a-t_r/2)}\int_{t_a-t_r/2}^{t_a+t_r/2}e^{i\omega (t-t_{min})}+e^{i\omega(t_a-t_r/2-t_{min})}\int_{t_a-t_r/2}^{t_a+t_r/2}\,dt\ g(t)e^{-i\omega (t-t_{min})}\big)\bigg].
\end{align}
In the limit $t_r\to 0$ we have \begin{align}
    \frac{d}{dt_r}\big(-\frac{1}{2}-n(\omega)\big)\big|C(t_a+t_r/2)-C(t_a-t_r/2)\big|^2_{t_r\to 0}=0.
    \end{align}
Next we have
\begin{align}
    \frac{d}{dt_r}\big(I(t_a+t_r/2)-I(t_a-t_r/2)\big)=\frac{i}{2m\omega}(1+2n_{\bar{\sigma}})\frac{1}{2}\bigg[g(t_a+t_r/2)\int_{t_{min}}^{t_a+t_r/2}\,dt\ g(t)\sin\big(\omega(t_a+t_r/2-t)\big)\\
    +g(t_a-t_r/2)\int_{t_{min}}^{t_a-t_r/2}\,dt\ g(t)\sin\big(\omega(t_a-t_r/2-t)\big)\bigg].
\end{align}
Also we have \begin{align}
    -i\frac{d}{dt_r}\int_{t_a-t_r/2}^{t_a+t_r/2}\,dt\ U(t)\hat{n}_{\bsig} = -iU(t_a)\hat{n}_{\bsig}.
\end{align}
The last term is given by
\begin{align}
        \frac{d}{dt_r}\frac{2ig(t_{min})}{\sqrt{2m\omega^3}}\hat{n}_f\text{Re}\bigg\{\big(C(t_a+t_r/2)-C(t_a-t_r/2)\big)\bigg\}\bigg|_{t_r\to 0}=\frac{2ig(t_{min})}{\sqrt{2m\omega^3}}\hat{n}_f\frac{1}{\sqrt{2m\omega}}g(t_a)\cos\left(\omega\left(t_a-t_{min}\right)\right).
\end{align}
So we've now calculated the derivatives of all the terms, so we may now write the full first derivative of $f$ in the $t_r\to 0$ limit: \begin{align}
f'(t_r\to 0)=i\mu-iU(t_a)\hat{n}_{\bsig} +\frac{i}{2m\omega}(1+2n_{\bar{\sigma}})g(t_a)\int_{t_{min}}^{t_a}\,dt\ g(t)\sin\big(\omega(t_a-t)\big)+\frac{ig(t_{min})}{m\omega^2}g(t_a)\cos\left(\omega( t_a-t_{min})\right).
\end{align}
\item \textbf{Second Derivatives}:\\ 
We have for the second derivative
\begin{align}
    &\frac{d^2}{dt_r^2}(-\frac{1}{2}-n(\omega))\big|C(t_a+t_r/2)-C(t_a-t_r/2)\big|^2=\nonumber\\
    &(-\frac{1}{2}-n(\omega))\frac{1}{4}\bigg[\frac{dg(t_a+t_r/2)}{dt_a}\big(e^{-i\omega(t_a+t_r/2-t_{min})}\int_{t_2}^{t_1}\,dt\ g(t)e^{i\omega( t-t_{min})}+\text{ c. c.}\big)\nonumber\\
    &+\frac{dg(t_a-t_r/2)}{dt_a}\big(e^{-i\omega(t_a-t_r/2-t_{min})}\int_{t_2}^{t_1}\,dt\ g(t)e^{i\omega (t-t_{min})}+ \text{ c.c.}\big)\nonumber\\
    &+g(t_a+t_r/2)\big(\big[-i\omega e^{-i\omega(t_a+t_r/2-t_{min})}\int_{t_2}^{t_1}\,dt\ g(t)e^{i\omega (t-t_{min})}-g(t_a-t_r/2)e^{i\omega t_r}+g(t_a+t_r/2)\big]+\text{ c.c.}\big)\nonumber\\
    &-g(t_a-t_r/2)\big(\big[i\omega e^{-i\omega(t_a-t_r/2-t_{min})}\int_{t_2}^{t_1}\,dt\ g(t)e^{i\omega (t-t_{min})}+g(t_a+t_r/2)e^{i\omega t_r}-g(t_a-t_r/2)\big]+\text{ c.c.}\big)\bigg].
\end{align}
So we again see in the limit $t_r\to 0$ \begin{align}
\frac{d^2}{dt_r^2}(-\frac{1}{2}-n(\omega))\big|C(t_a+t_r/2)-C(t_a-t_r/2)\big|^2\big|_{t_r\to 0}=0.
\end{align}
Next we have
\begin{align}
    &\frac{d^2}{dt_r^2}\big(I(t_a+t_r/2)-I(t_a-t_r/2)\big)=\frac{i}{4m\omega}(1+2n_{\bar{\sigma}})\bigg\{\frac{1}{2}\frac{dg(t_a+t_r/2)}{dt_r}\int_{t_{min}}^{t_1}\,dt\ g(t)\sin\big(\omega(t_a+t_r/2-t)\big)\nonumber\\
    &-\frac{1}{2}\frac{dg(t_a-t_r/2)}{dt_r}\int_{t_{min}}^{t_2}\,dt\ g(t)\sin\big(\omega(t_a-t_r/2-t)\big)+\frac{\omega}{2}g(t_a+t_r/2)\int_{t_{min}}^{t_1}\,dt\ g(t)\cos\big(\omega(t_a+t_r/2-t)\big)\nonumber\\
    &-\frac{\omega}{2}g(t_a-t_r/2)\int_{t_{min}}^{t_2}\,dt\ g(t)\cos\big(\omega(t_a-t_r/2-t)\big)\bigg\}
\end{align}
which in the limit $t_r\to 0$ becomes\begin{align}
  \frac{d^2}{dt_r^2}\big(I(t_a+t_r/2)-I(t_a-t_r/2)\big)\big|_{t_r\to 0}=0.
  \end{align}
  Next,
  \begin{align}
  &\frac{d^2}{dt_r^2}\big(C(t_1)-C(t_2)\big)=\frac{1}{4\sqrt{2m\omega}}\big(\frac{dg(t_a+t_r/2)}{dt_r}e^{i\omega(t_a+t_r/2-t_{min})}+i\omega g(t_a+t_r/2)e^{i\omega(t_a+t_r/2-t_{min})}\nonumber\\
  &-\frac{dg(t_a+t_r/2)}{dt_r}e^{i\omega(t_a-t_r/2-t_{min})}-i\omega g(t_a-t_r/2)e^{i\omega(t_a-t_r/2-t_{min})}\big),
  \end{align}
 and so the limit $t_r\to 0$ gives \begin{align}
 \frac{d^2}{dt_r^2}\big(C(t_1)-C(t_2)\big)\big|_{t_r\to 0}=0.
 \end{align}
 Finally, we have \begin{align}
     &-i\frac{d^2}{dt_r^2}\int_{t_a-t_r/2}^{t_a+t_r/2}\,dt\ U(t)\hat{n}_{\bsig}=-\frac{i}{2}\frac{d}{dt_r}\left[U(t_a+t_r/2)+U(t_a-t_r/2)\right]\nonumber\\
     &=\frac{-i}{4}\hat{n}_{\bsig}\left[\frac{dU(t_a+t_r/2)}{dt_a}-\frac{dU(t_a-t_r/2)}{dt_a}\right]
 \end{align}
 which vanishes in the limit $t_r\to 0$.
 So we now have the second derivative of $f$ in the $t_r\to 0$ limit, \begin{align}
 f''(t_r\to 0)=0.
 \end{align}
 \item \textbf{Third Derivatives}:\\
 For the third derivative we have
 \begin{align}
     &\frac{d^3}{dt_r^3}(-\frac{1}{2}-n(\omega))\big|C(t_a+t_r/2)-C(t_a-t_r/2)\big|^2 =\nonumber\\
     &(-\frac{1}{2}-n(\omega))\frac{1}{8}\bigg[\frac{d^2g(t_a+t_r/2)}{dt_a^2}\big(e^{-i\omega(t_a+t_r/2-t_{min})}\int_{t_2}^{t_1}\,dt\ g(t)e^{i\omega(t-t_{min})}+ \text{ c.c.}\big)\nonumber\\
     &-\frac{d^2g(t_a-t_r/2)}{dt_a^2}\big(e^{-i\omega(t_a-t_r/2-t_{min})}\int_{t_2}^{t_1}\,dt\ g(t)e^{i\omega(t-t_{min})}+ \text{ c.c.}\big)\nonumber\\
     &+2\frac{dg(t_a+t_r/2)}{dt_a}\big(\big[-i\omega e^{-i\omega(t_a+t_r/2-t_{min})}\int_{t_2}^{t_1}\,dt\ g(t)e^{i\omega (t-t_{min})}-g(t_a-t_r/2)e^{i\omega t_r}+g(t_a+t_r/2)\big]+\text{ c.c.}\big)\nonumber\\
     &+2\frac{dg(t_a-t_r/2)}{dt_a}\big(\big[i\omega e^{-i\omega(t_a-t_r/2-t_{min})}\int_{t_2}^{t_1}\,dt\ g(t)e^{i\omega (t-t_{min})}+g(t_a+t_r/2)e^{i\omega t_r}-g(t_a-t_r/2)\big]+\text{ c.c.}\big)\nonumber\\
     &+2g(t_a+t_r/2)\bigg(\big(-\frac{\omega^2}{2}e^{-i\omega(t_a+t_r/2-t_{min})}\int_{t_2}^{t_1}\,dt\ g(t)e^{i\omega (t-t_{min})}-\frac{i\omega}{2}g(t_a+t_r/2)+\frac{i\omega}{2}g(t_a-t_r/2)e^{-i\omega t_r}\nonumber\\
     &+\frac{1}{2}\frac{dg(t_a-t_r/2)}{dt_a}e^{-i\omega t_r}+i\omega g(t_a-t_r/2)e^{-i\omega t_r}+\frac{1}{2}\frac{dg(t_a+t_r/2)}{dt_a}\big]+\text{ c.c.}\big)\bigg)\nonumber\\
     &-2g(t_a-t_r/2)\bigg(\big[\frac{\omega^2}{2}e^{-i\omega(t_a-t_r/2-t_{min})}\int_{t_2}^{t_1}\,dt\ g(t)e^{i\omega (t-t_{min})}+\frac{i\omega}{2}\big(g(t_a+t_r/2)e^{i\omega t_r}-g(t_a-t_r/2)\big)\nonumber\\
     &+\frac{1}{2}\frac{dg(t_a+t_r/2)}{dt_a}e^{i\omega t_r}_i\omega g(t_a+t_r/2)e^{i\omega t_r}+\frac{1}{2}\frac{dg(t_a-t_r/2)}{dt_a}\big]+\text{ c.c.}\bigg)\bigg]
 \end{align}
 and hence in the limit $t_r\to 0$ we have
 \begin{align}
 \frac{d^3}{dt_r^3}(-\frac{1}{2}-n(\omega))\big|C(t_a+t_r/2)-C(t_a-t_r/2)\big|^2\big|_{t_r\to 0}=0.
 \end{align}
 Next,
 \begin{align}
     &\frac{d^3}{dt_r^3}\big(I(t_a+t_r/2)-I(t_a-t_r/2)\big)\nonumber\\&=\frac{i}{8m\omega}(1+2n_{\bar{\sigma}})\bigg\{\frac{1}{2}\frac{d^2g(t_a+t_r/2)}{dt^2_a}\int_{t_{min}}^{t_a+t_r/2}\,dt\ g(t)\sin\big(\omega(t_a+t_r/2-t)\big)\nonumber\\
     &+\frac{1}{2}\frac{d^2g(t_a-t_r/2)}{dt_a^2}\int_{t_{min}}^{t_a-t_r/2}\,dt\ g(t)\sin\big(\omega(t_a-t_r/2-t)\big)\nonumber\\
     &+\omega\frac{dg(t_a+t_r/2)}{dt_a}\int_{t_{min}}^{t_a+t_r/2}\,dt\ g(t)\cos\big(\omega(t_a+t_r/2-t)\big)\nonumber\\
     &+\omega\frac{dg(t_a-t_r/2)}{dt_a}\int_{t_{min}}^{t_a-t_r/2}\,dt\ g(t)\cos\big(\omega(t_a-t_r/2-t)\big)\nonumber\\
     &+\frac{\omega}{2}g^2(t_a+t_r/2)+\frac{\omega}{2}g^2(t_a-t_r/2)\nonumber\\
     &-\frac{\omega^2}{2}g(t_a-t_r/2)\int_{t_{min}}^{t_a+t_r/2}\,dt\ g(t)\sin\big(\omega(t_a+t_r/2-t)\big)\nonumber\\
     &-\frac{\omega^2}{2}g(t_a-t_r/2)\int_{t_{min}}^{t_a-t_r/2}\,dt\ g(t)\sin\big(\omega(t_a-t_r/2-t)\big)\bigg\}
 \end{align}
 so that in the limit $t_r\to 0$ we have:
 \begin{align}
 &\frac{d^3}{dt_r^3}\big(I(t_a+t_r/2)-I(t_a-t_r/2)\big)\big|_{t_r\to 0}
 =\frac{i}{8m\omega}(1+2n_{\bar{\sigma}})\bigg\{\frac{d^2g(t_a)}{dt_a^2}\int_{t_{min}}^{t_a}\,dt\ g(t)\sin\big(\omega(t_a-t)\big)\nonumber\\
 &+2\omega\frac{dg(t_a)}{dt_a}\int_{t_{min}}^{t_a}\,dt\ g(t)\cos\big(\omega(t_a-t)\big)+ \omega g^2(t_a)\nonumber\\
&-\omega g(t_a)\int_{t_{min}}^{t_a}\,dt\ g(t)\sin\big(\omega(t_a-t)\big).
 \end{align}
Next, we have \begin{align}
     &\frac{d^3}{dt_r^3}\big(C(t_a+t_r/2)-C(t_a-t_r/2)\big)=\frac{1}{8\sqrt{2m\omega}}\bigg(\frac{d^2g(t_a+t_r/2)}{dt_a^2}e^{i\omega(t_a+t_r/2-t_{min})}\nonumber\\
     &+2i\omega\frac{dg(t_a+t_r/2)}{dt_a}e^{i\omega(t_a+t_r/2-t_{min})}+2i\omega\frac{dg(t_a-t_r/2)}{dt_a}e^{i\omega(t_a-t_r/2-t_{min})}\nonumber\\
     &+\frac{d^2g(t_a-t_r/2)}{dt_a^2}e^{i\omega(t_a-t_r/2)}-\omega^2g(t_a+t_r/2)e^{i\omega(t_a+t_r/2)}-\omega^2g(t_a-t_r/2)e^{i\omega(t_a-t_r/2)}\bigg).
 \end{align}
In the limit $t_r\to 0$ we have \begin{align}
&\frac{2ig(t_{min})n_f}{\sqrt{2m\omega^3}}\text{Re}\bigg\{ \frac{d^3}{dt_r^3}\big(C(t_a+t_r/2)-C(t_a-t_r/2)\big)\bigg\}\big|_{t_r\to 0} \nonumber\\
&=\frac{i g(t_{min})n_f}{4m\omega^2}\bigg\{\frac{d^2g(t_a)}{dt_a^2}\cos\left(\omega \left(t_a-t_{min}\right)\right)-2\omega\frac{dg(t_a)}{dt_a}\sin\left(\omega\left(t_a-t_{min}\right)\right)-\omega^2g(t_a)\cos\left(\omega \left(t_a-t_{min}\right)\right)\bigg\}.
\end{align}
Finally we have \begin{align}
    -i\frac{d^3}{dt_r^3}\int_{t_a-t_r/2}^{t_a+t_r/2}\,dt\ U(t)\hat{n}_{\bsig}\vert_{t_r\to 0}=-i\frac{1}{4}\frac{d^2U(t_a)}{dt_a^2}n_{\bsig}.
\end{align}

So finally
we have for the third derivative of $f$ in the $t_r\to 0$ limit, \begin{align}
&f'''(t_r\to 0) \nonumber\\
&=\frac{i g(t_{min})n_f}{4m\omega^2}\bigg\{ \frac{d^2g(t_a)}{dt_a^2}\cos\left(\omega\left( t_a-t_{min} \right)\right)-2\omega\frac{dg(t_a)}{dt_a}\sin\left(\omega\left( t_a-t_{min}\right)\right)-\omega^2g(t_a)\cos\left(\omega\left(t_a-t_{min}\right)\right)\bigg\}\nonumber\\
&+\frac{i}{8m\omega}(1+2n_{\bar{\sigma}})\bigg\{\frac{d^2g(t_a)}{dt_a^2}\int_{t_{min}}^{t_a}\,dt\ g(t)\sin\big(\omega(t_a-t)\big)+2\omega\frac{dg(t_a)}{dt_a}\int_{t_{min}}^{t_a}\,dt\ g(t)\cos\big(\omega(t_a-t)\big)+ \omega g^2(t_a)
\nonumber\\& -\omega g(t_a)\int_{t_{min}}^{t_a}\,dt\ g(t)\sin\big(\omega(t_a-t)\big)-\frac{i}{4}\frac{d^2U(t_a)}{dt_a^2}n_{\bsig}\bigg\}.
 \end{align}
 We've now completely determined all the derivatives in the $t_r$ to zero limit: $f', f'', f'', u', u'', u'''$. It remains to put everything together. We are now ready to plug into the formulas we derived above (Eqn's \ref{first},\ref{second},\ref{third}), and evaluate. 
\end{enumerate}
\subsection{Evaluating the Moments}

The zeroth moment, $\mu^{R0}_{at,\sigma}(t_a)=1$ is trivially satisfied.
\subsubsection{First Moment}
From equation \ref{first}, we have \begin{align}
    &F'(t_r\to 0)=f'(t_r\to 0) +u'(t_r\to 0)(1-2n_{\sigma})\nonumber\\
    &= i\mu-iU(t_a)n_{\bar{\sigma}} +\frac{i}{2m\omega}(1+2n_{\bar{\sigma}})g(t_a)\int_{t_{min}}^{t_a}\,dt\ g(t)\sin\big(\omega(t_a-t)\big)+\frac{ig(t_{min})}{m\omega^2}g(t_a)\cos(\omega\left( t_a-t_{min}\right))\nonumber\\&-(1-2n_{\sigma})\frac{i}{2m\omega}g(t_a)\text{Re}\bigg\{ie^{-i\omega t_a}\int_{t_{min}}^{t_a}\,dt\ g(t)e^{i\omega t}\bigg\}\nonumber\\
    &=i\big(\mu-U(t_a)n_{\bar{\sigma}}-g(t_a)x(t_a)\big).
\end{align}
Plugging into the moment definition and evaluating the trace we see \begin{align}
&\mu_{at,\sigma}^{R1}(t_a)=-\mu+U(t_a)\langle n_{\bar{\sigma}}\rangle+g(t_a)\langle x(t_a)\rangle
\end{align}
and the first moment sum rule is satisfied.
\subsubsection{Second Moment}
From Eq. (\ref{second}) we have, noting $u''(t_r\to 0)=f''(t_r\to 0)=0$, \begin{align}
     &F''(t_r\to 0)= f''+f'^2+2f'u'(1-2n_{\sigma})+u'^2+u''(1-2n_{\sigma})\nonumber\\
     &=\big(f'+u'(1-2n_{\sigma})\big)^2\nonumber\\
     &=i^2\big(\mu-U(t_a)n_{\bar{\sigma}}-g(t_a)x(t_a)\big)^2.
\end{align}
When we evaluate the trace, we obtain \begin{align}
    &\mu_{\sigma,at}^{R2}(t_a)=(\mu-U(t_a)\langle n_{\bar{\sigma}}\rangle-g(t_a)\langle x(t_a)\rangle)^2+g^2(t_a)\big(\langle x^2(t_a)\rangle -\langle x(t_a)\rangle ^2\big)\nonumber\\&+U^2(t_a)\left(\langle n_{\bar{\sigma}}^2\rangle-\langle n_{\bar{\sigma}}\rangle^2\right)+g(t_a)U(t_a)\left(\langle n_{\bar{\sigma}}x(t_a)\rangle-\langle n_{\bar{\sigma}}\rangle\langle x(t_a)\rangle\right)
\end{align}
and the second moment sum rule is satisfied.
\subsubsection{Third Moment}
Finally for the third moment, we have from Eq. (\ref{third}), recalling $u''=f''=0$\begin{align}
     &F'''(t_r\to 0)=(f'''+3f'f''+f'^3)+3(f''+f'^2)u'(1-2n_{\sigma})+3f'u'^2+3f'u''(1-2n_{\sigma})\nonumber\\
     &+3u'u''+(u'^3+u''')(1-2n_{\sigma})\nonumber\\
     &=\big(f'+u'(1-2n_{\sigma})\big)^3+f'''+u'''(1-2n_{\sigma}).
\end{align}

The cubed term is simply the first moment cubed. The remaining term we need to evaluate is \begin{align}
   &f'''+ u'''(1-2n_{\sigma})=\frac{ig^2(t_a)\omega}{2m\omega}+\frac{ig^2(t_a)\omega}{4m\omega}\big(n_{\bar{\sigma}}-3n_{\sigma}\big)-\frac{i}{4}\frac{d^2U(t_a)}{dt_a^2}n_{\bsig}\nonumber\\
   &+\frac{i}{4m\omega}(n_{\sigma}+n_{\bar{\sigma}})\bigg\{\frac{d^2g(t_a)}{dt_a^2}\text{Re}\big\{ie^{-i\omega t_a}\int_{t_{min}}^{t_a}\,dt\ g(t)e^{i\omega t}\big\}+2\omega\frac{dg(t_a)}{dt_a}\int_{t_{min}}^{t_a}\,dt\ g(t)\cos\big(\omega(t-t_a)\big)\nonumber\\
   &-\omega^2g(t_a)\text{Re}\big\{ie^{-i\omega t_a}\int_{t_{min}}^{t_a}\,dt\ g(t) e^{i\omega t}\big\}\bigg\} \nonumber\\
   &+\frac{ig(t_{min})}{4m\omega^2}(n_{\sigma}+n_{\bar{\sigma}})\bigg\{\frac{d^2g(t_a)}{dt_a^2}\cos\left(\omega \left(t_a-t_{min}\right)\right)-2\omega\frac{dg(t_a)}{dt_a}\sin\left(\omega \left(t_a-t_{min}\right)\right)-\omega^2\cos\left(\omega\left( t_a-t_{min}\right)\right)\bigg\}.
\end{align}
Simplifying and evaluating the trace, this becomes \begin{align}
    &f'''+(1-2n_{\sigma})u'''=\frac{i}{4}g(t_a)\omega^2\langle x(t_a)\rangle -\frac{i}{4}\frac{d^2g(t_a)}{dt_a^2}\langle x(t_a)\rangle -\frac{i}{2}\frac{dg(t_a)}{dt_a}\frac{d\langle x(t_a)\rangle}{dt_a}\nonumber\\
    &+i\frac{g^2(t_a)\omega}{2m\omega}+i\frac{g^2(t_a)\omega)}{4m\omega}\big(\langle n_{\bar{\sigma}}\rangle -3\langle n_{\sigma}\rangle \big)-\frac{i}{4}\frac{d^2U(t_a)}{dt_a^2}n_{\bsig}.
\end{align}
Note $\langle n_{\sigma}\rangle =\langle n_{\bar{\sigma}}\rangle = \frac{1}{2}\langle n_{\sigma}+n_{\bsig}\rangle$, so the third moment becomes \begin{align}
&\mu_{\sigma,at}^{R3}=-\big(\mu-U(t_a)\langle n_{\bsig}\rangle-g(t_a)\langle x(t_a)\rangle \big)^3+g^3(t_a)\big(\langle x^3(t_a)\rangle -\langle x(t_a)\rangle^3\big)+U^3(t_a)\big(\langle n_{\bsig}\rangle-\langle n_{\bsig}\rangle^3\big)\nonumber\\
&-3\mu g^2(t_a)\big(\langle x^2(t_a)\rangle -\langle x(t_a)\rangle^2\big)-3\mu U^2(t_a)\big(\langle n_{\bsig}\rangle-\langle n_{\bsig}\rangle^2\big)+3U(t_a)g^2(t_a)\big(\langle n_{\bsig}x^2(t_a)\rangle-\langle n_{\bsig}\rangle\langle x(t_a)\rangle^2 \big)\nonumber\\
    &-6\mu g(t_a)U(t_a)\big(\langle n_{\bsig}x(t_a)\rangle-\langle n_{\bsig}\rangle\langle x(t_a)\rangle\big)+3U^2(t_a)g(t_a)\big(\langle n_{\bar{\sigma}}x(t_a)\rangle-\langle n_{\bar{\sigma}}\rangle^2\langle x(t_a)\rangle\big)\nonumber\\
&+\frac{1}{4}g(t_a)\omega^2\langle x(t_a)\rangle -\frac{1}{4}\text{Re}\left\{\frac{d^2g(t_a)}{dt_a^2}\langle x(t_a)\rangle+\frac{d^2U(t_a)}{dt_a^2}\langle n_{\bsig}\rangle\right\}+\frac{g^2(t_a)\omega}{2m\omega}\big(1-\frac{1}{2}\langle n_{\sigma}+n_{\bsig}\rangle\big)-\frac{1}{2}\frac{dg(t_a)}{dt_a}\frac{d\langle x(t_a)\rangle}{dt_a}.
\end{align}
So we have now analytically confirmed that the causal Green's function for the non-equilibrium Holstein-Hubbard model presented in the main text exactly satisfies the first four moment sum rules as given in Ref. 34.